\documentclass[eqsecnum,aps,preprint]{revtex4}
\usepackage{epsfig,graphicx}
\begin{document}

\title{
Centrality Dependence of Thermal Parameters Deduced from Hadron Multiplicities 
in Au + Au Collisions at $\sqrt{s_{NN}} = 130\ GeV$ }

\author{
{\sc J. Cleymans$^a$, B. K\"ampfer$^b$, M. Kaneta$^c$, S. Wheaton$^a$, N. Xu$^d$} }

\address{
$^a$ Department of Physics, University of Cape Town,
Rondebosch 7701, Cape Town, South Africa\\
$^b$ Institut f\"ur Kern- und Hadronenphysik,
Forschungszentrum Rossendorf, PF 510119, D-01314 Dresden, Germany\\
$^c$ RIKEN BNL Research Center, Brookhaven National Laboratory, Upton, New York 11973\\
$^d$ Lawrence Berkeley National Laboratory, Berkeley, California 94720\\
}

\begin{abstract}
We analyse the centrality dependence of thermal parameters deduced from hadron multiplicities in
Au + Au collisions at $\sqrt{s_{NN}} = 130\ GeV$. While the chemical freeze-out temperature and
chemical potentials are found to be roughly centrality-independent, 
the strangeness saturation factor $\gamma_S$
increases with participant number towards unity, supporting the assumption of
equilibrium freeze-out conditions in central collisions.
\end{abstract}

\maketitle

\section{Introduction} 

Statistical-thermal models (cf.~\cite{review} for recent surveys and references therein) 
have enjoyed remarkable success in describing hadron multiplicities 
observed in both heavy-ion and elementary collisions over a wide range of energies
\cite{heavyions,abundancesa,abundancesb,cor,further_refs}.
The final state multiplicities are reproduced in these models with very
few parameters. The prominent ones are the chemical freeze-out temperature $T$ and baryon chemical
potential $\mu_B$. There are further chemical potentials to be considered below. A compilation of
the freeze-out parameters over the accessible range of energies reveals a continuous curve 
in the $T-\mu_B$ plane which
may be characterized by an energy per hadron of the order of $1\ GeV$ \cite{abundancesb}.
While the occurrence of such a universal
freeze-out curve is interesting in itself and is useful for extrapolations and interpolations as
well, its particular meaning could be the approximate coincidence with the confinement border
line, at least at high energies, as conjectured in \cite{C12}.

We do not touch here on the disputed question as to why also the multiplicities 
of the rare multi-strange
hadrons seem to obey chemical equilibrium (see \cite{equilibrium_discussion}). Instead, we focus
on one possible indicator of incomplete equilibrium, the strangeness saturation factor $\gamma_s$.
This factor has been introduced \cite{gamma_s}
to account for an apparent under-saturation of
strange hadrons. Various previous analyses \cite{previous_analyzes}
considered such a factor as necessary to
accomplish a satisfactory description of data. 
For SPS energies, $\sqrt{s_{NN}} = 6 \cdots 20\ GeV$,
it has been shown \cite{we1} that $\gamma_s$ grows with increasing beam energy and centrality. 
First attempts \cite{we2}, based on very restricted data sets, find also at RHIC energies
a preliminary indication of rising $\gamma_s$ with increasing centrality.
It is the
subject of this paper to investigate in detail the dependence of the thermal parameters 
including $\gamma_s$
on the centrality in collisions of Au+Au at $\sqrt{s_{NN}} = 130\ GeV$. 
By now there is a sufficiently
large basis of published data at this energy to allow a thorough analysis.

Our paper is organized as follows. In section II we recall the formulation of our employed
statistical-thermal model. The analysed data are listed in section III. Our 
fit procedures are described in section IV and results are presented
in section V. The summary and conclusions can be found in section VI.

\section{Statistical-thermal model} 

For suitably large systems at RHIC collider energies, hadron multiplicities are analysed by
employing the grand-canonical partition function ${\cal Z} (V, T, \mu_i) = \mbox{Tr} \left[
\exp\{- (\hat H - \sum_i \mu_i Q_i) / T \} \right]$, where $\hat H$ is the statistical operator of
the system, $T$ denotes the temperature, and $\mu_i$ and $Q_i$ are respectively the chemical
potentials and corresponding conserved charges in the system. To be specific, we use here the
baryon charge (corresponding to $\mu_B$) and the strangeness charge (corresponding to $\mu_S$) as
unconstrained quantities. (There are other possible chemical potentials, e.g.,
$\mu_Q$ as electric charge potential. However, 
we put $\mu_Q = 0$ with the reasoning explained below.)

The primordial particle numbers are accordingly,
\begin{equation}
N_i^{\rm prim} = Vg_i \int \frac{d^3 p}{(2\pi)^3} \, dm_i \, \left[ \gamma_s^{-|S_i|}
\mbox{e}^{\frac{E_i - \mu_i Q_i}{T}} \pm 1 \right]^{-1} \mbox{BW} (m_i),
\label{eq.1}
\end{equation}
where we include phenomenologically the strangeness saturation factor $\gamma_s$ (with $|S_i|$ the
total number of strange and anti-strange quarks in hadron species $i$) to account for the
possibility of incomplete equilibration in this sector; $E_i = \sqrt{p^2 + m_i^2}$ and
$\mbox{BW}(m_i)$ is the Breit-Wigner distribution (to be replaced by a $\delta$-function for
stable hadrons) to be integrated from thresholds with appropriate widths $\Gamma_i$
(in practice the interval $[max \{threshold, m_i - 2 \Gamma_i\},m_i + 2 \Gamma_i]$ 
is sufficient). 
The fiducial volume $V$ drops in particle ratios. $g_i$ is the degeneracy factor
of particle species $i$ with vacuum mass $m_i$.
(The study of the effect of in-medium masses (cf.\ \cite{ff}) is interesting
because it might shed light on the chiral properties of the medium
created in heavy-ion collisions at RHIC.)
Eq.~(\ref{eq.1}) simplifies in the Boltzmann approximation (i.e., discarding the
spin-statistics factor $\pm 1$) and when neglecting the energy distribution of resonances.  
We use here the full expression Eq.~(\ref{eq.1}) and include all hadron states with $u, d, s$
quarks and corresponding anti-quarks 
up to 1.7 or $2.6\ GeV$ \footnote{Actually we use two independent codes with 
different upper limits of the employed hadronic mass spectrum. 
The corresponding differences in results are tiny. One of the codes
is described in S. Wheaton, J. Cleymans, hep-ph/0407174.} 
with masses and total widths according to the particle data group listing \cite{PDG}. 
The final particle
numbers to be compared with experiment are $N_i = N_i^{\rm prim} + \sum_j \mbox{Br}^{j \to i}
N_j^{\rm prim}$ due to decays of unstable particles with branching ratios
$\mbox{Br}^{j \to i}$.

Originally, such a description was thought to be justified for multiplicities measured over the
entire phase-space, since many dynamical effects cancel out in ratios of hadron yields 
\cite{cor}.
At sufficiently high energy, such as at RHIC, however, ratios of mid-rapidity yields are also
found to be well described by the statistical-thermal model 
\cite{review,abundancesa,we2,NuXuKaneta,Becanttini}.

\section{The Data}

In contrast to SPS energies, where both mid-rapidity and fully phase space integrated data are at
our disposal (see \cite{we2} for a comparison), 
at RHIC energy of $\sqrt{s_{NN}} = 130\ GeV$
sufficiently many data of identified hadrons are available only at mid-rapidity. 
The data analysed here were
compiled from PHENIX, PHOBOS, BRAHMS and STAR experimental results. These experiments differ in
their cuts on the data. 
It turns out that the results for the deduced thermal freeze-out parameters depend in some
cases sensitively on the employed data sets. 
Therefore, for definiteness
Table \ref{Table:Data} 
lists the ratios used as a function of collision
centrality after recalculation to a common centrality binning as in \cite{NuXuKaneta}. 
Subscript ''(1)'' denotes yields corrected for weak feed-down, 
while yields labelled by ''(2)'' include a contribution from weak decays.
Three centrality bins were selected with participant numbers $63.5 \pm 8.4$ (peripheral), $210.5
\pm8.4$ (mid-central) and $317.0 \pm8.2$ (central).

\begin{table}[!]
\caption{Ratios of hadrons in $Au+Au$ collisions at $\sqrt{s_{NN}} = 130\ GeV$
from various RHIC experiments.}\label{Table:Data}
\begin{tabular}{|c|c|c|c|c|}\hline
Ratio & Experiment & Central & Mid-Central & Peripheral\\ \hline
$\pi^-_{(2)}$/$\pi^+_{(2)}$             & BRAHMS \cite{d23}& 0.990$\pm$0.100 &  & \\
                                        & PHENIX \cite{d24,d25} & 0.960$\pm$0.177 & 0.920$\pm$0.170 & 0.933$\pm$0.172\\
                                        & PHOBOS \cite{d26}& 1.000$\pm$0.022 &  & \\
                    & STAR  \cite{d21} & 1.000$\pm$0.073 & 1.000$\pm$0.073 & 1.000 $\pm$ 0.073 \\
$K^+_{(2)}/K^-_{(2)}$                   & PHENIX \cite{d24,d25} & 1.152$\pm$0.240 & 1.292$\pm$0.268 & 1.322$\pm$0.284 \\
                                        & PHOBOS \cite{d26}& 1.099$\pm$0.111 &  & \\
                                        & STAR   \cite{d27}& 1.109$\pm$0.022 & 1.105$\pm$0.036 & 1.120$\pm$0.040 \\
$\bar{p}_{(1)}/p_{(1)}$                 & PHENIX \cite{d24,d25}& 0.680$\pm$0.149 & 0.671$\pm$0.142 & 0.717$\pm$0.157 \\
$\bar{p}_{(2)}/p_{(2)}$                 & BRAHMS \cite{d23}& 0.650$\pm$0.092 &  & \\
                                        & PHOBOS \cite{d26}& 0.600$\pm$0.072 &  & \\
                                        & STAR   \cite{d30}& 0.714$\pm$0.050 & 0.724$\pm$0.050 & 0.764$\pm$0.053 \\
$\bar{\Lambda}_{(1)}/\Lambda_{(1)}$     & PHENIX \cite{d25}& 0.750$\pm$0.180 & 0.798$\pm$0.197 & 0.795$\pm$0.197  \\
$\bar{\Lambda}_{(2)}/\Lambda_{(2)}$     & STAR   \cite{d22}& 0.719$\pm$0.090 & 0.739$\pm$0.092 & 0.744$\pm$0.100 \\
$\bar{\Xi}^+_{(2)}/\Xi^-_{(2)}$         & STAR   \cite{d32,d33}& 0.840$\pm$0.053 & 0.822$\pm$0.114 & 0.815$\pm$0.096 \\
$\bar{\Omega}^+/\Omega^-$               & STAR   \cite{d33,d34}& 1.062$\pm$0.410 & &  \\
$K^-_{(2)}/\pi^-_{(2)}$                 & PHENIX \cite{d24,d25}& 0.151$\pm$0.030 & 0.134$\pm$0.027 & 0.116$\pm$0.023 \\
                                        & STAR   \cite{d21,d27}& 0.151$\pm$0.022 & 0.147$\pm$0.022 & 0.130$\pm$0.019 \\
$K^0_S/\pi^-_{(2)}$                     & STAR   \cite{d21,d27}& 0.134$\pm$0.022 & 0.131$\pm$0.022 & 0.108$\pm$0.018 \\
$\bar{p}_{(1)}/\pi^-_{(2)}$             & PHENIX \cite{d24,d25}& 0.049$\pm$0.010 & 0.047$\pm$0.010 & 0.045$\pm$0.009 \\
$\bar{p}_{(2)}/\pi^-_{(2)}$             & STAR   \cite{d21,d29}& 0.069$\pm$0.019 & 0.067$\pm$0.019 & 0.067$\pm$0.019 \\
$\Lambda_{(1)}/\pi^-_{(2)}$             & STAR   \cite{d21,d22}& 0.043$\pm$0.008 & 0.043$\pm$0.008 & 0.039$\pm$0.007 \\
$\Lambda_{(2)}/\pi^-_{(2)}$             & PHENIX \cite{d24,d25}& 0.072$\pm$0.017 & 0.068$\pm$0.016 & 0.074$\pm$0.017  \\
$<K^{*0}>/\pi^-_{(2)}$                  & STAR   \cite{d21,d28}& 0.039$\pm$0.011 & &   \\
$\phi/\pi^-_{(2)}$                      & STAR   \cite{d21,d31}& 0.022$\pm$0.003 & 0.021$\pm$0.004 & 0.022$\pm$0.004 \\
$\Xi^-_{(2)}/\pi^-_{(2)}$               & STAR   \cite{d21,d32,d33}& 0.0093$\pm$0.0012 & 0.0072$\pm$0.0011 & 0.0060$\pm$0.0008 \\
$\Omega^-/\pi^-_{(2)}$                  & STAR   \cite{d21,d32,d33}& 0.0014$\pm$0.0004 & &   \\
\hline
\end{tabular}
\end{table}

\section{The Fit Procedures}

According to our propositions, $T, \mu_B, \mu_S$ and $\gamma_s$
are unconstrained fit parameters.
For each centrality class, four fits were performed: \\
Fit I: all available ratios listed in Table I included;\\ 
Fit II: only ratios of $\pi$, $K$, $p$, $\Lambda$ and $\Xi$ included;\\
Fit III: only ratios of $\pi$, $K$, $p$ and $\Lambda$ included; \\
Fit IV: only ratios of $\pi$, $K$ and $p$ included.\\ 
In this way, the effect of the various multiplicities on the
thermal parameters are investigated. (Various other fitting procedures are
considered in \cite{NuXuKaneta}.)

The final ratios compared with experiment include both a primordial
and a decay contribution. As mentioned above, 
in some cases the experimental data have been corrected for weak decays.
However, where such corrections have not been made,
the influence of weak decays is included in the following way:\\
\begin{tabular}{rl}
$\pi^\pm_{(2)}$ :& includes 50\% of pions from decays of $\Omega$, $\Xi$, $\Sigma$, 
$\Lambda$ and their anti-particles\\
$K^\pm_{(2)}$ :& includes 100\% of kaons from $\phi$ decay\\
               & and 50\% of kaons from $\Omega$ and $\bar{\Omega}$ decay\\
$K^0_s$ :& includes 100\% of $K^0_s$ from $\phi$ decay\\
$p_{(2)}$ and $\bar{p}_{(2)}$ :& includes 100\% of protons from the decay of $\Lambda$, $\Sigma^0$ 
and their anti-particles\\
$\Lambda_{(2)}$ and $\bar{\Lambda}_{(2)}$:& includes 100\% of the contribution 
from the decay of $\Xi$, $\Omega$\\
              &  and their anti-particles\\
$\Xi^-_{(2)}$ and $\bar{\Xi}^+_{(2)}$ :
& includes 50\% of the contribution from $\Omega$ and $\bar{\Omega}$ decay\\
\end{tabular}\\

As seen in Table I, in some cases  the same ratios are at our
disposal from various experiments. 
We include these multiply given ratios in our analysis as separate data points to be fit. 

\section{Results}

\subsection{Centrality dependence of thermal parameters}

\begin{figure}
\includegraphics[width=7cm]{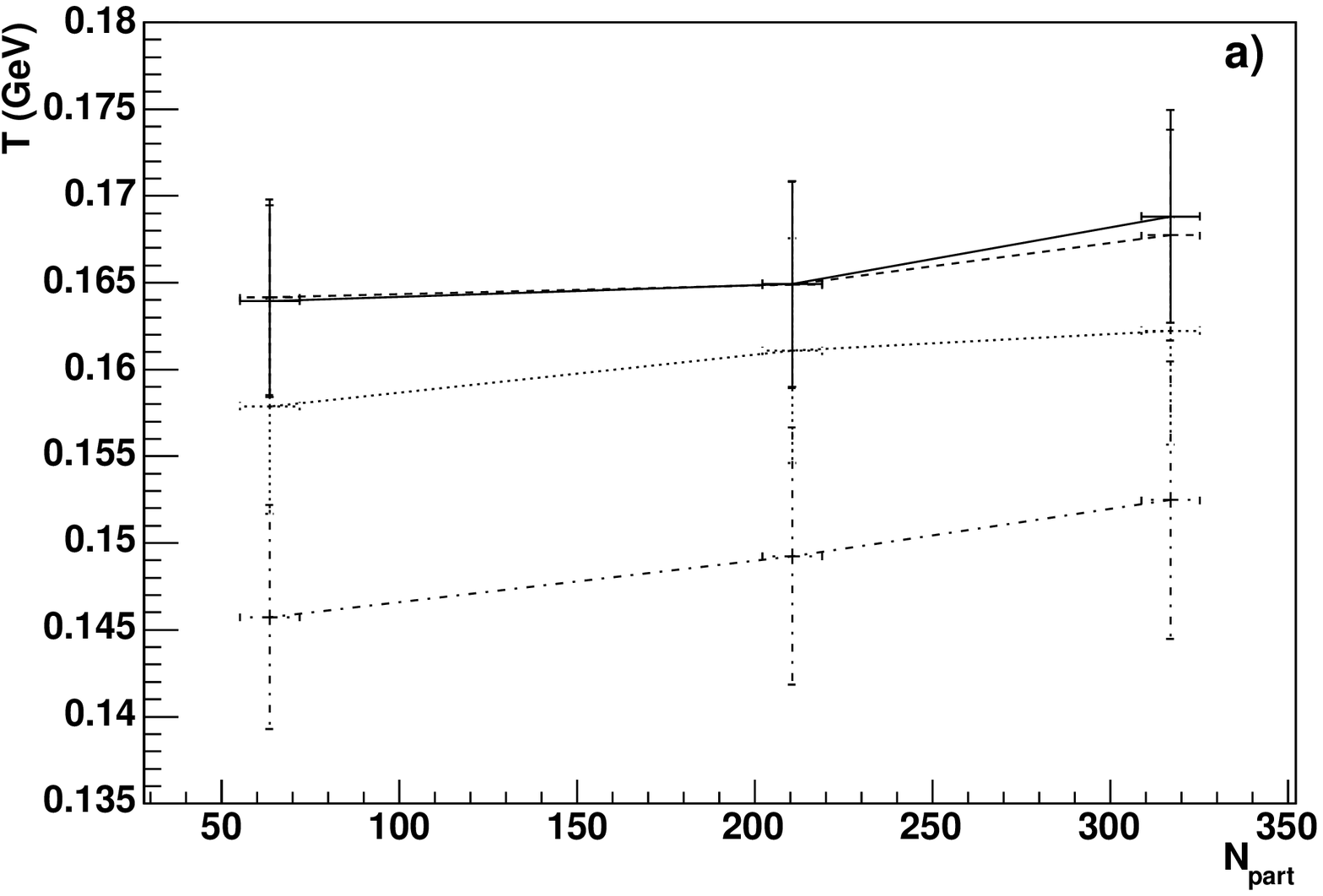}
\includegraphics[width=7cm]{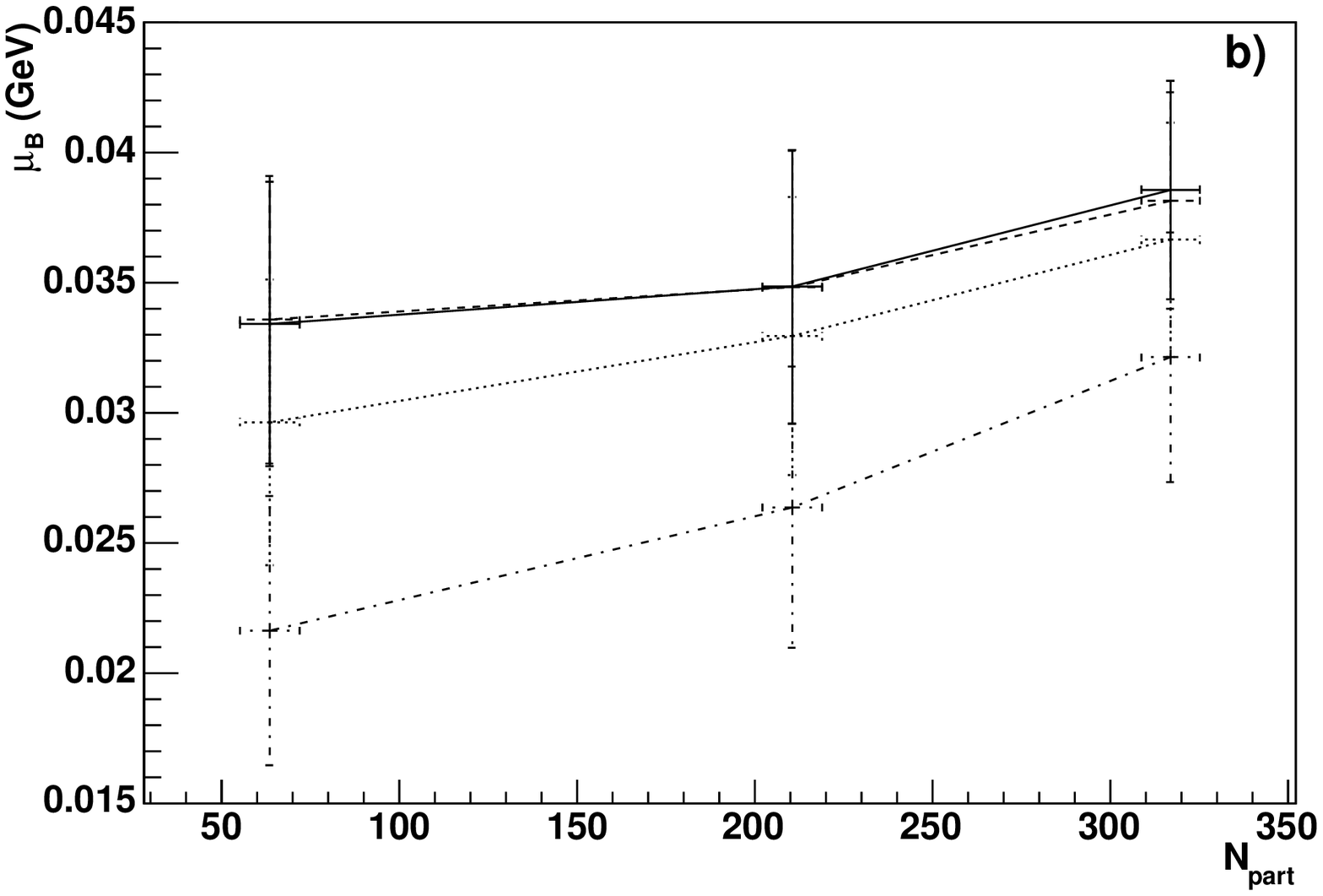}\\
\includegraphics[width=7cm]{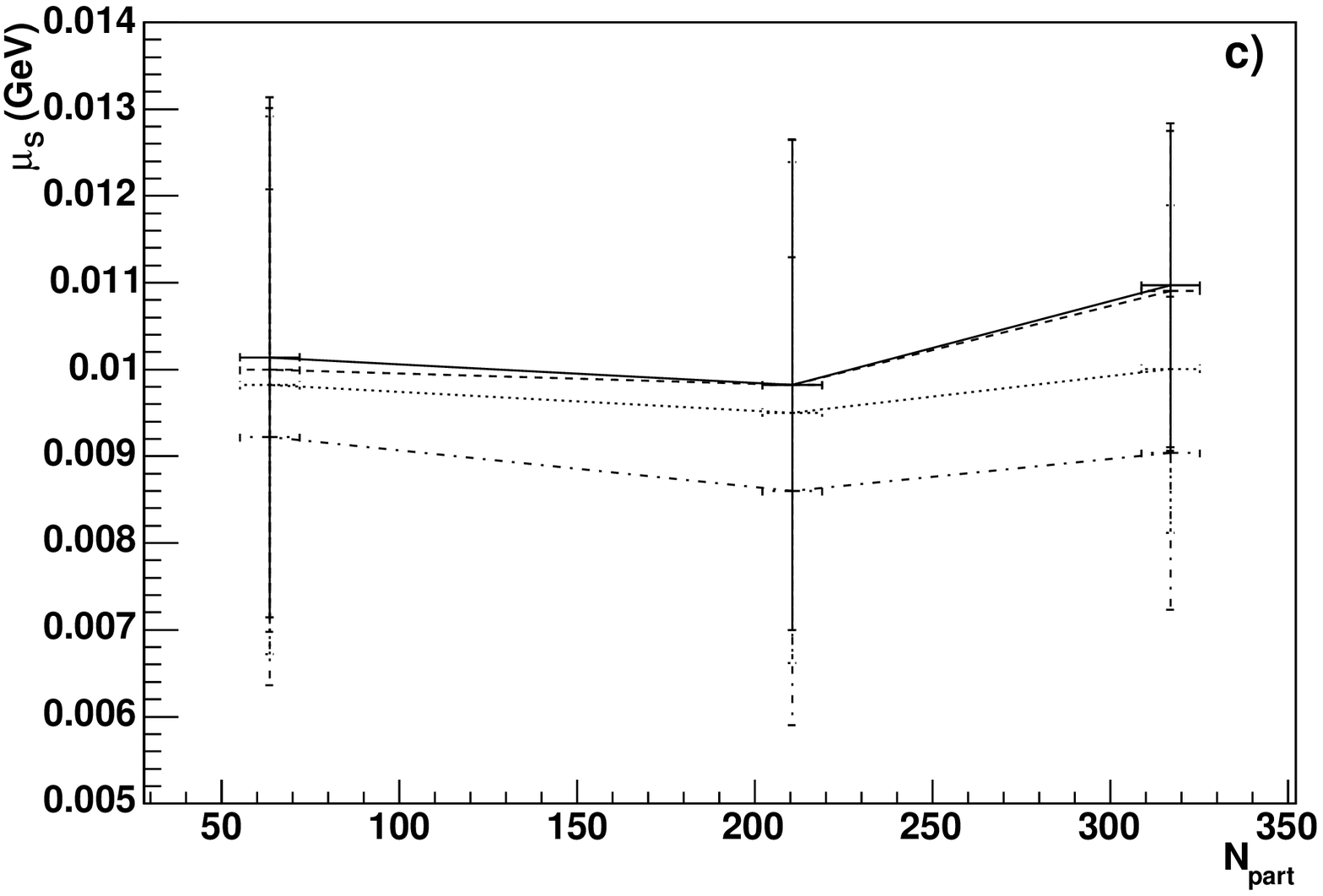}
\includegraphics[width=7cm]{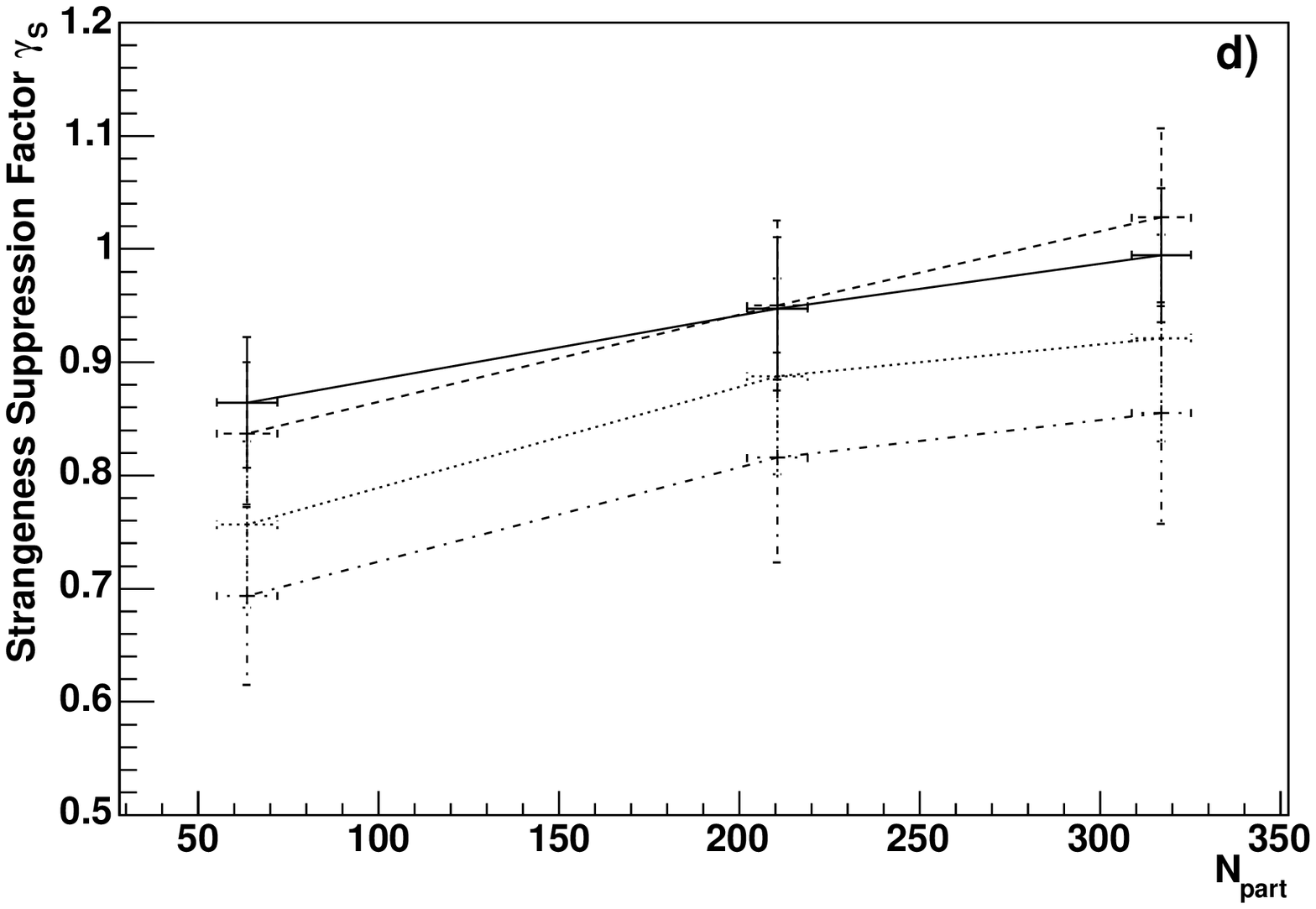}
\caption{The thermal parameters (a: temperature $T$, b: baryon chemical potential $\mu_B$,
c: strangeness chemical potential $\mu_S$, d: strangeness saturation factor $\gamma_s$)
as extracted from the experimental data for the various fit types 
(Fit I: solid, Fit II: dashed, Fit III: dotted, Fit IV: dash-dotted curves).
}\label{Results:Graph}
\end{figure}

The global results of the fits are displayed in Fig.~\ref{Results:Graph}.
As is evident from Fig.~\ref{Results:Graph}a, the
chemical freeze-out temperature is roughly constant at around $165\ MeV$ in Fit I. 
While the strangeness chemical potential is fairly  flat at $\mu_S\approx10\ MeV$ 
(with 10\% variation, Fig.~\ref{Results:Graph}c),
the baryon chemical potential (Fig.~\ref{Results:Graph}b)
increases slightly with participant
number in the range $33.4\ MeV$ to $38.5\ MeV$ (a 14\% increase) in Fit I. Most
striking, however, and the main result of our analysis,
is the increase in strangeness saturation (Fig.~\ref{Results:Graph}d) 
with increasing centrality for all fit types. This confirms previous arguing \cite{we2,NuXuKaneta}.
On a quantitative level, an analysis based on too few hadron species delivers
non-reliable results, in particular also for $\gamma_s$ (see Fig.~\ref{Results:Graph}d, Fit IV).
Fits I and II, in contrast, deliver consistent results.
Since we consider the phenomenologically introduced parameter $\gamma_s$
as one possible indicator of deviations from equilibrium, the conclusion is that
in central collisions equilibrium conditions for describing the chemical freeze-out
are appropriate, while more peripheral collisions point to some off-equilibrium effects. 

\begin{figure}
\includegraphics[width=4.5cm]{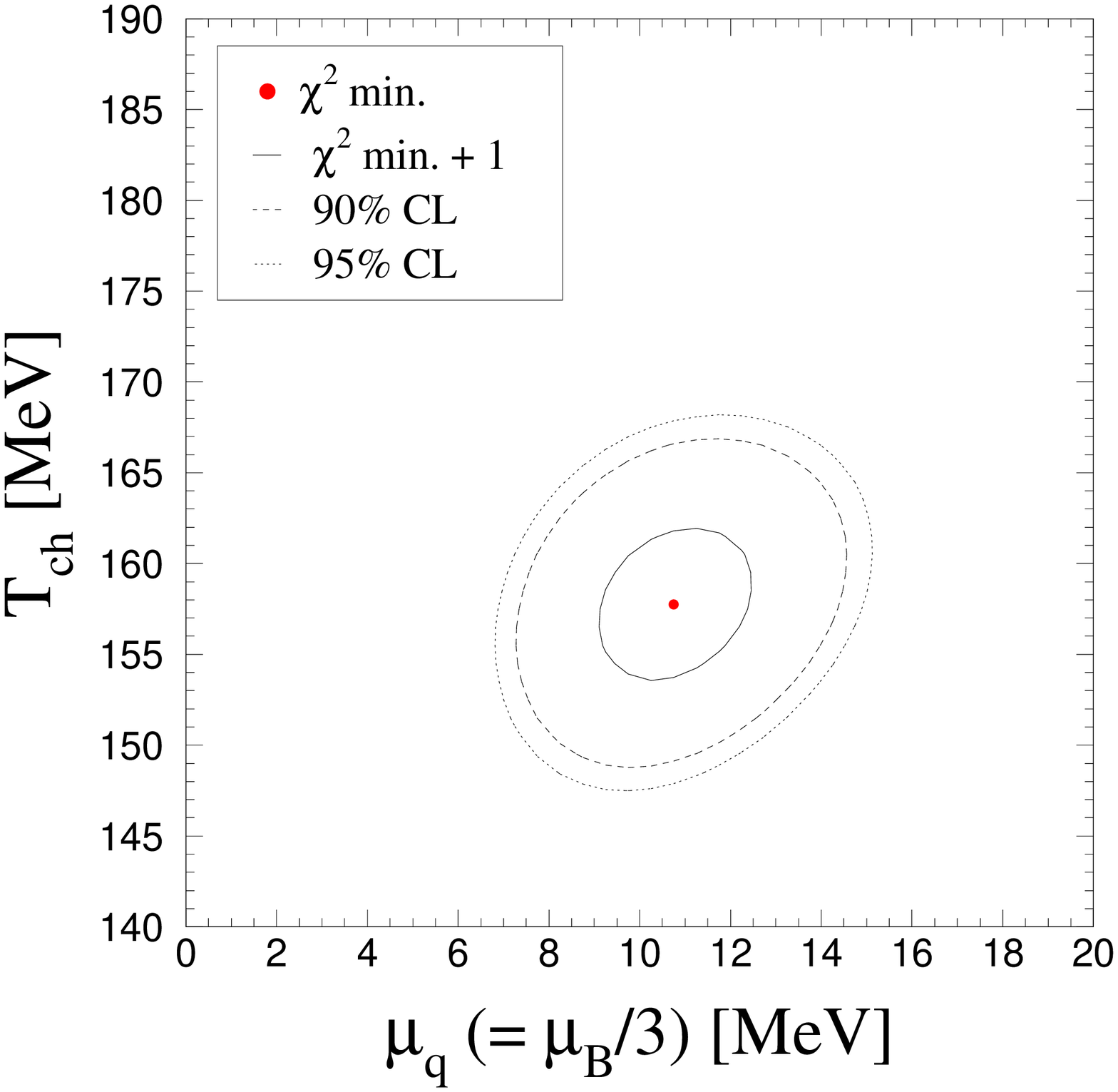}
\includegraphics[width=4.5cm]{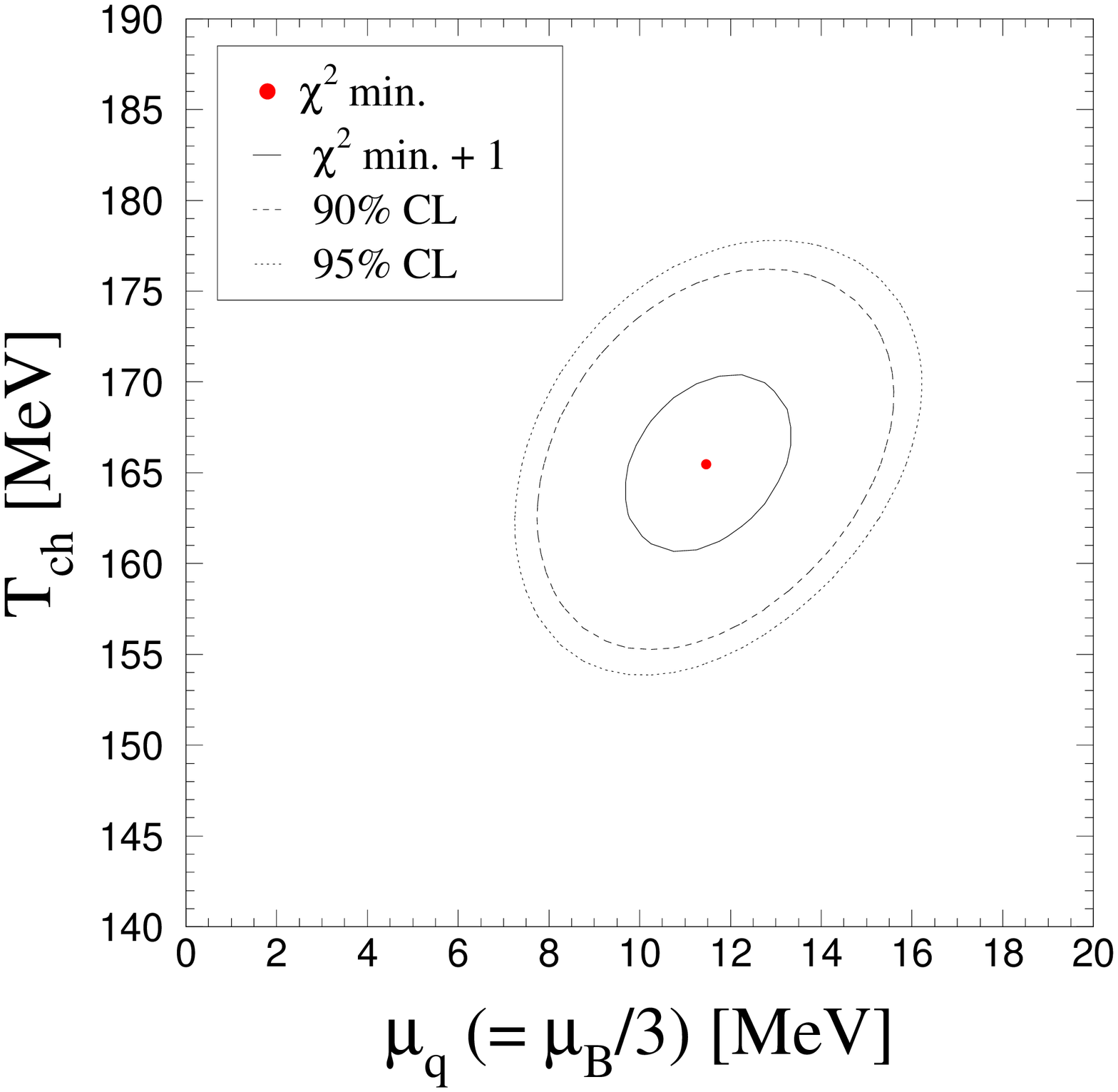}
\includegraphics[width=4.5cm]{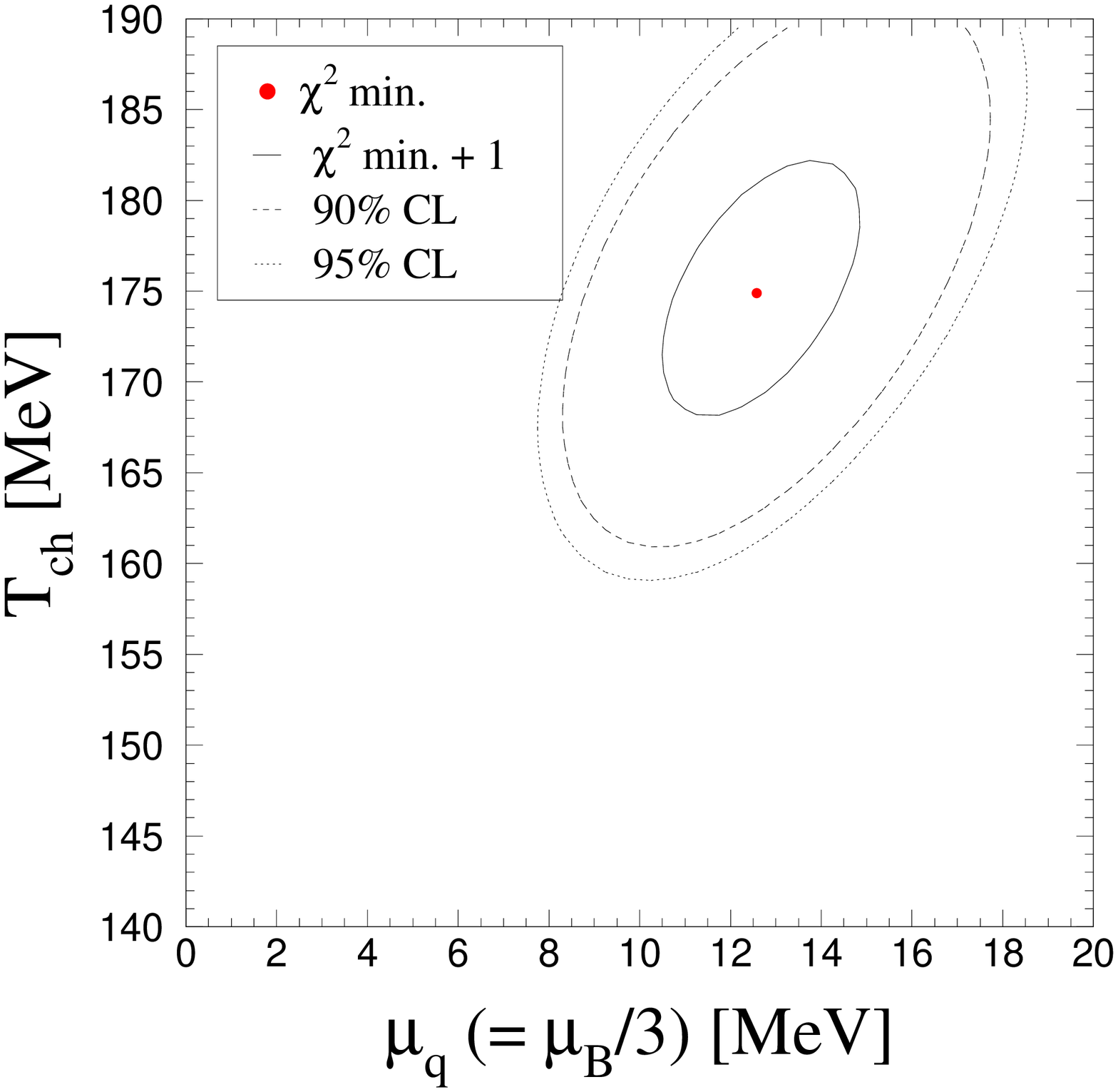}
\caption{$\chi^2$ contours from Fit I of peripheral data. $\gamma_s = 1$ (0.5)
in left (right) panel. In the middle panel, $\gamma_s = 0.85$ being the optimum choice.}
\label{contour}
\end{figure}

To highlight the role of the parameter $\gamma_s$ on the quality of the data fits, we
exhibit in Fig.~\ref{contour} the $\chi^2$ contours for fixed values of $\gamma_s$
in the $T - \mu$ plane. For definiteness, only results for Fit I and peripheral
collisions are shown. We have selected $\gamma_s = 1$ (representing the assumption
of full equilibrium), 0.85 (the optimum choice)
and 0.5 (as rather extreme choice), respectively. 
The corresponding $\chi^2/d.o.f$ are 24.1/12, 17.2/11 and 62.0/12.
The figure evidences that lowering $\gamma_s$ noticeably shifts $T$ up, while
the up-shift of $\mu_B$ is less pronounced. The change in $\chi^2/d.o.f$ quantifies
the relevance of the parameter $\gamma_s$ for the fits.

\subsection{Detailed comparison with data}

After discussing the global trends, it is worthwhile to look into details.
Instead of displaying the usual comparison of model results with data in one plot
we consider each ratio individually to exhibit the centrality systematics. 

\subsubsection{Particle-anti--particle ratios}

\begin{figure}
\includegraphics[width=7cm]{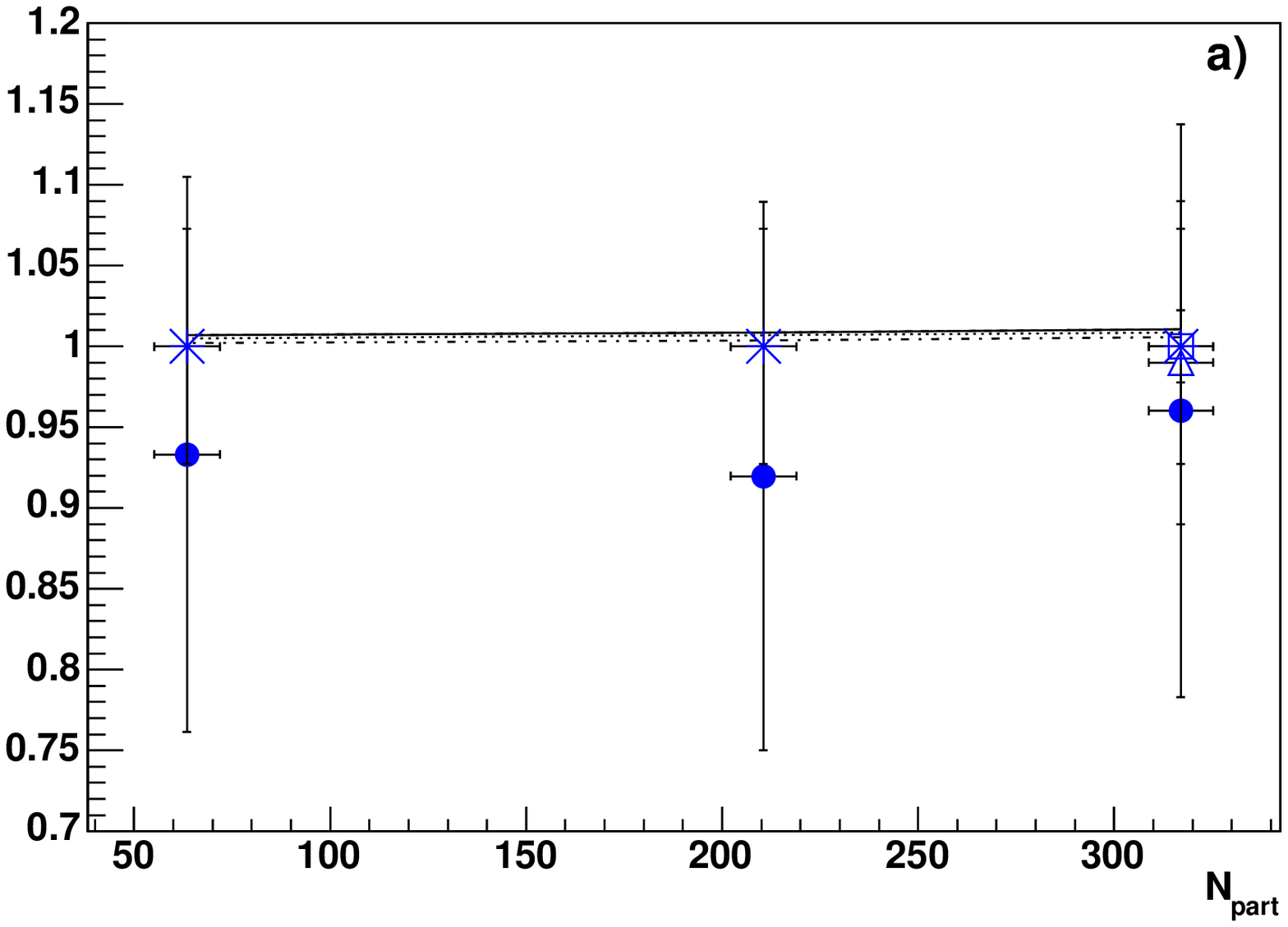}
\includegraphics[width=7cm]{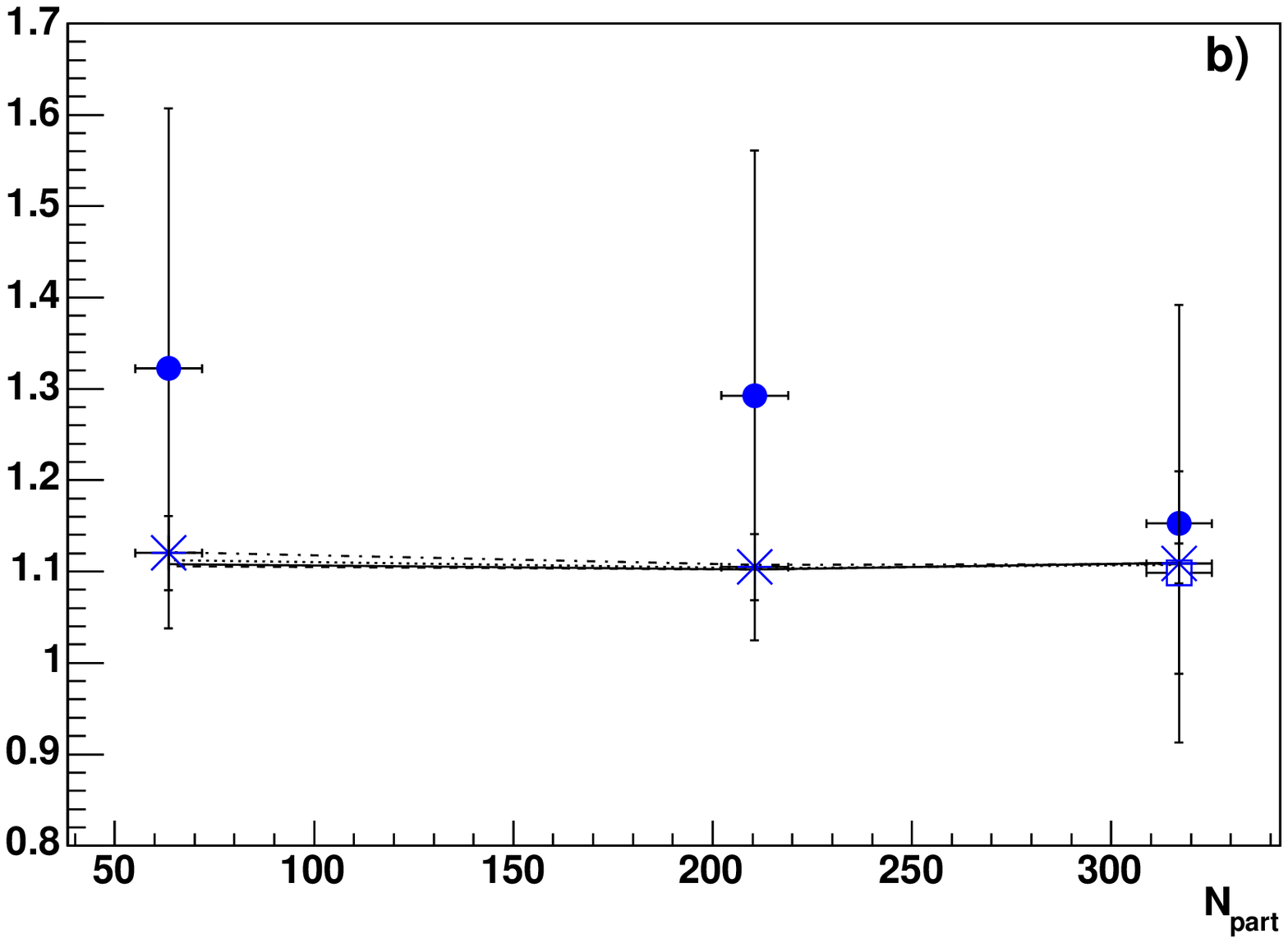}
\includegraphics[width=7cm]{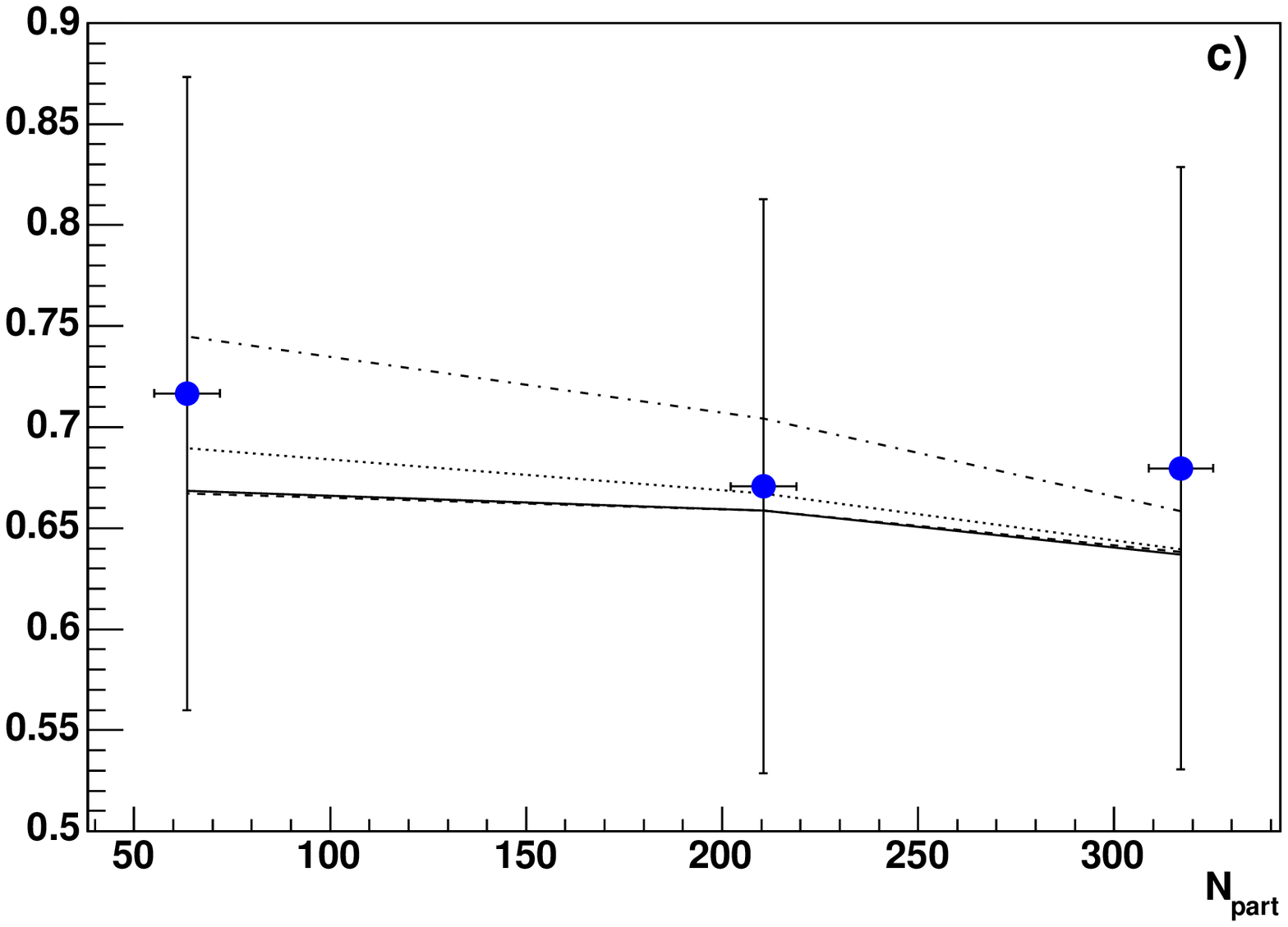}
\includegraphics[width=7cm]{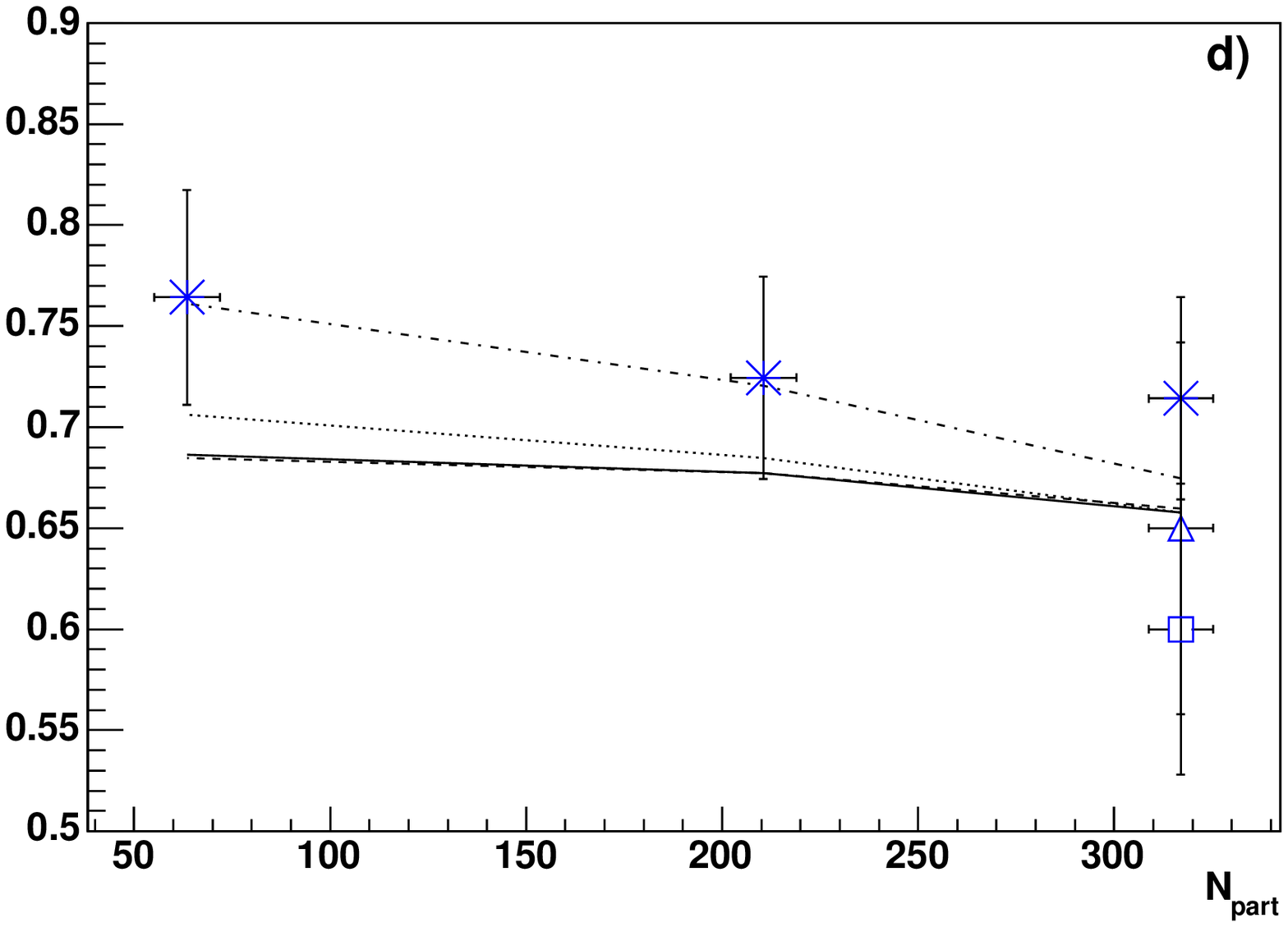}
\includegraphics[width=7cm]{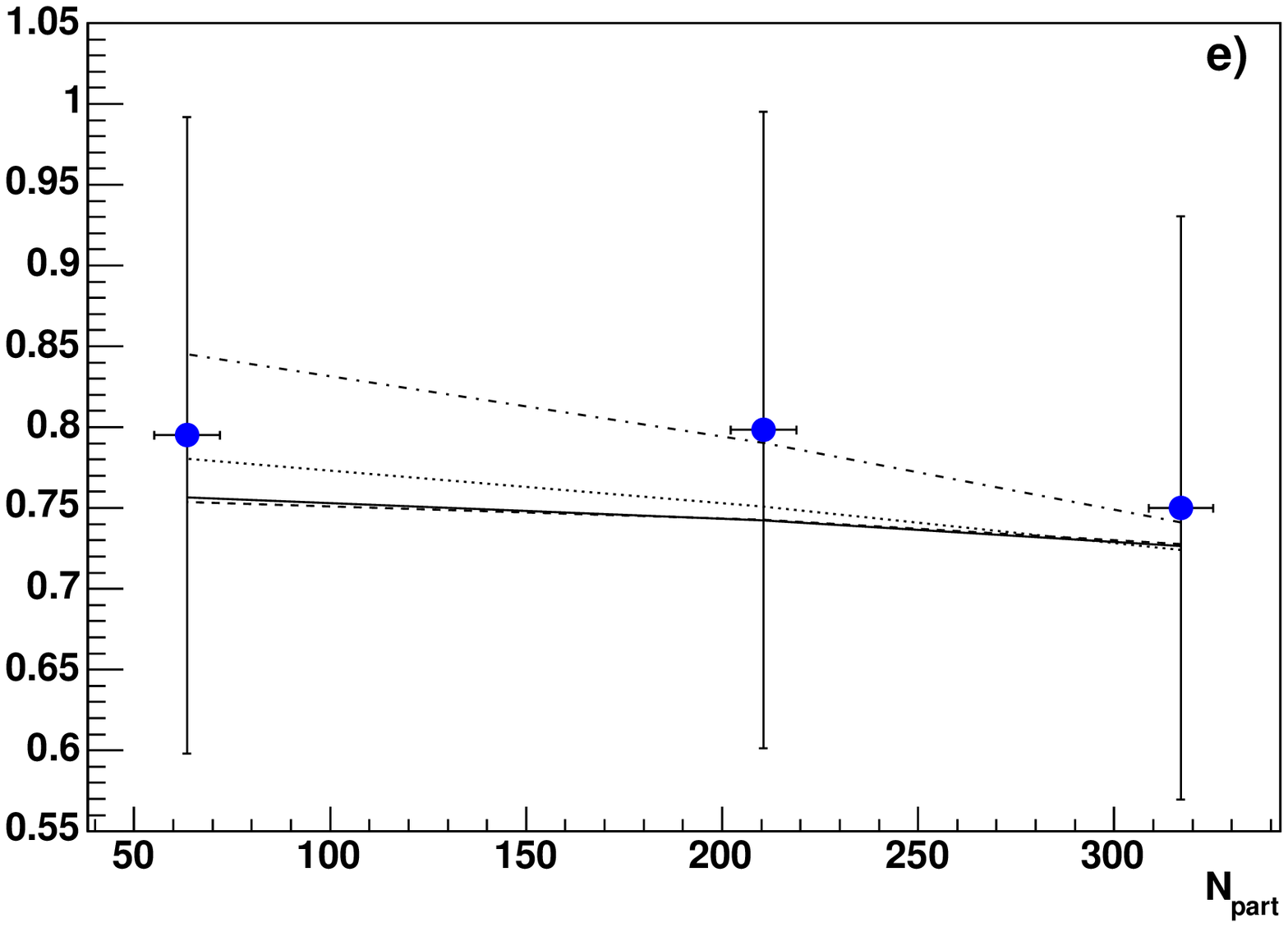}
\includegraphics[width=7cm]{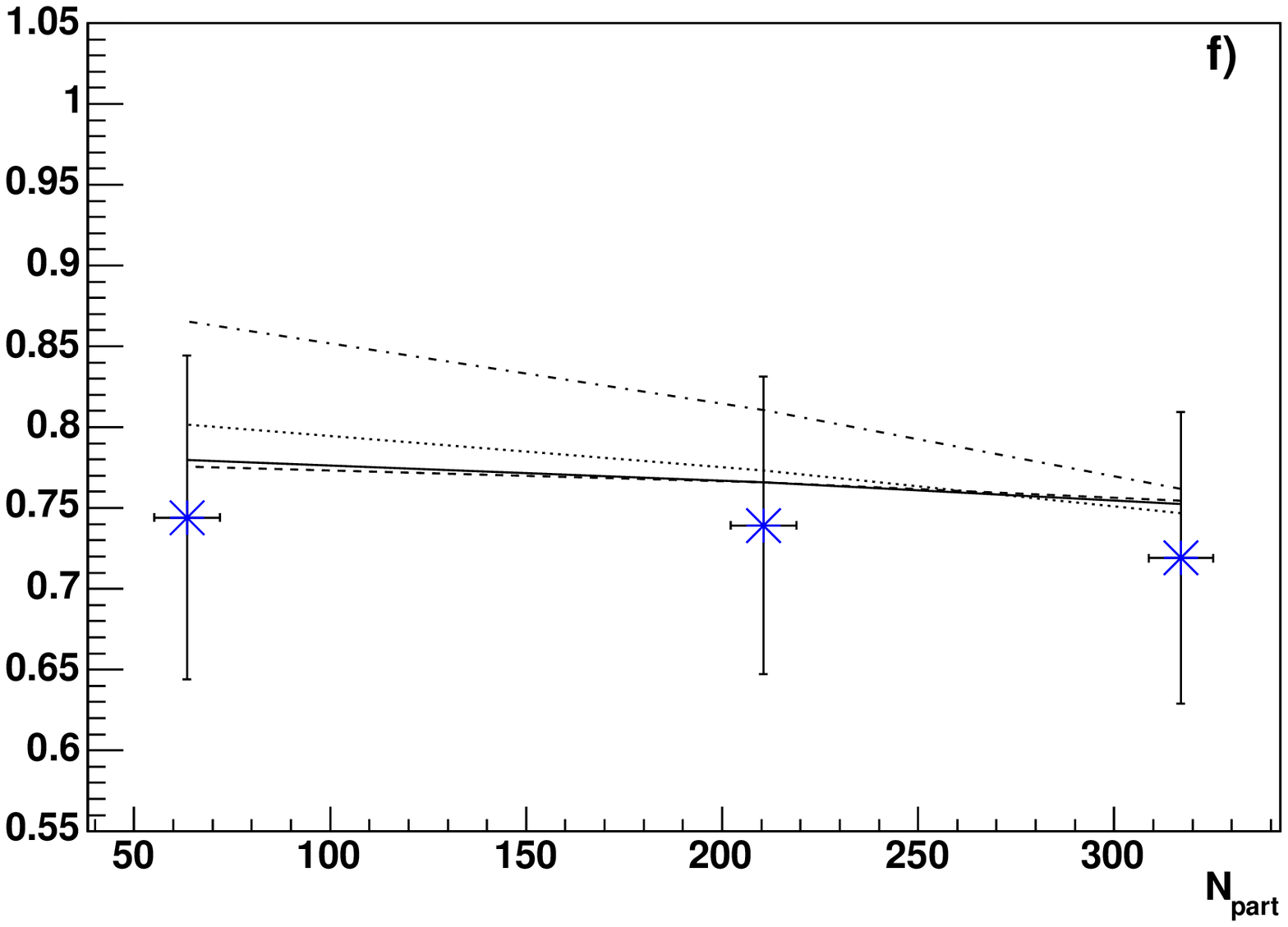}
\includegraphics[width=7cm]{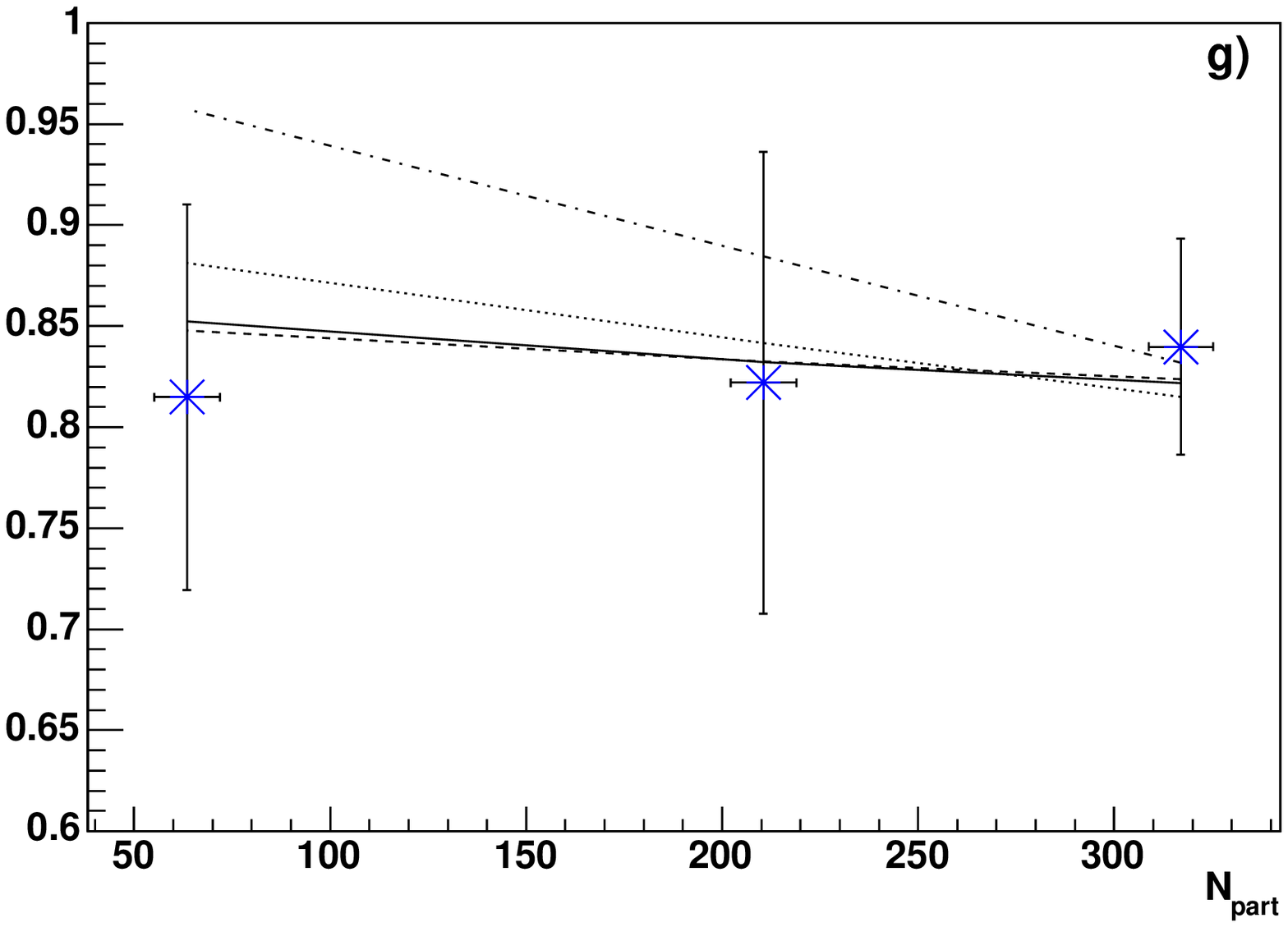}
\includegraphics[width=7cm]{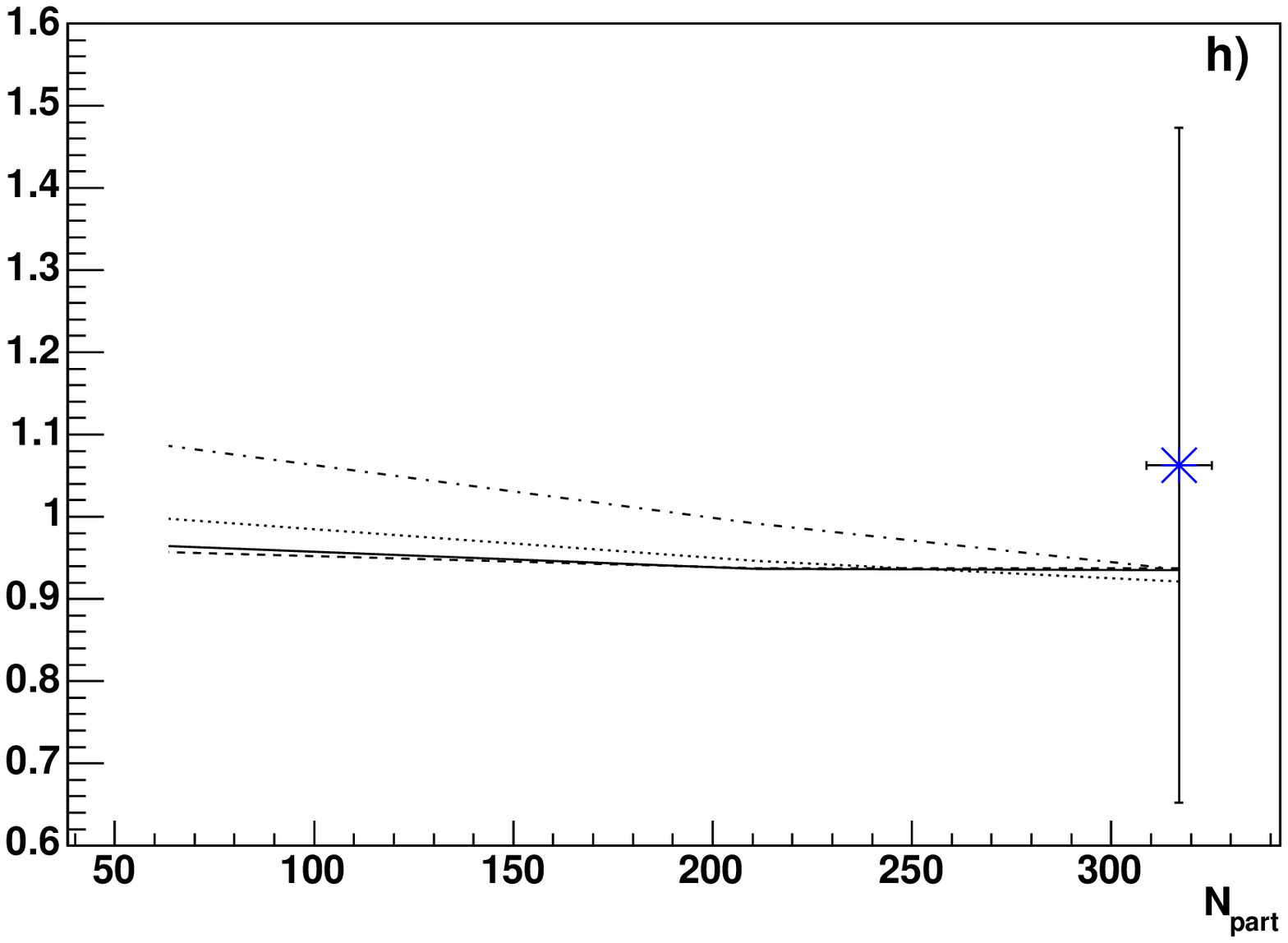}
\caption{Comparison between experiment and model results for
particle--anti-particle ratios 
(a: $\pi^-_{(2)}/\pi^+_{(2)}$, 
b: $K^+_{(2)}/K^-_{(2)}$, 
c: $\bar{p}_{(1)}/p_{(1)}$, 
d: $\bar{p}_{(2)}/p_{(2)}$, 
e: $\bar{\Lambda}_{(1)}/\Lambda_{(1)}$, 
f: $\bar{\Lambda}_{(2)}/\Lambda_{(2)}$, 
g: $\bar{\Xi}^+_{(2)}/\Xi^-_{(2)}$, 
h: $\bar{\Omega}^+/\Omega^-$).
Line codes as in Fig.~\ref{Results:Graph}, 
experimental results: STAR: stars, PHENIX: circles,
PHOBOS: squares, BRAHMS: triangles, cf.\ Table I.
}
\label{Results:ModelPAP}
\end{figure}

Fig.~\ref{Results:ModelPAP}
summarizes such a comparison for the three centrality classes for the 
particle-anti--particle ratios.
The data used from Table I are shown together with our model results.

\begin{figure}
\includegraphics[width=7cm]{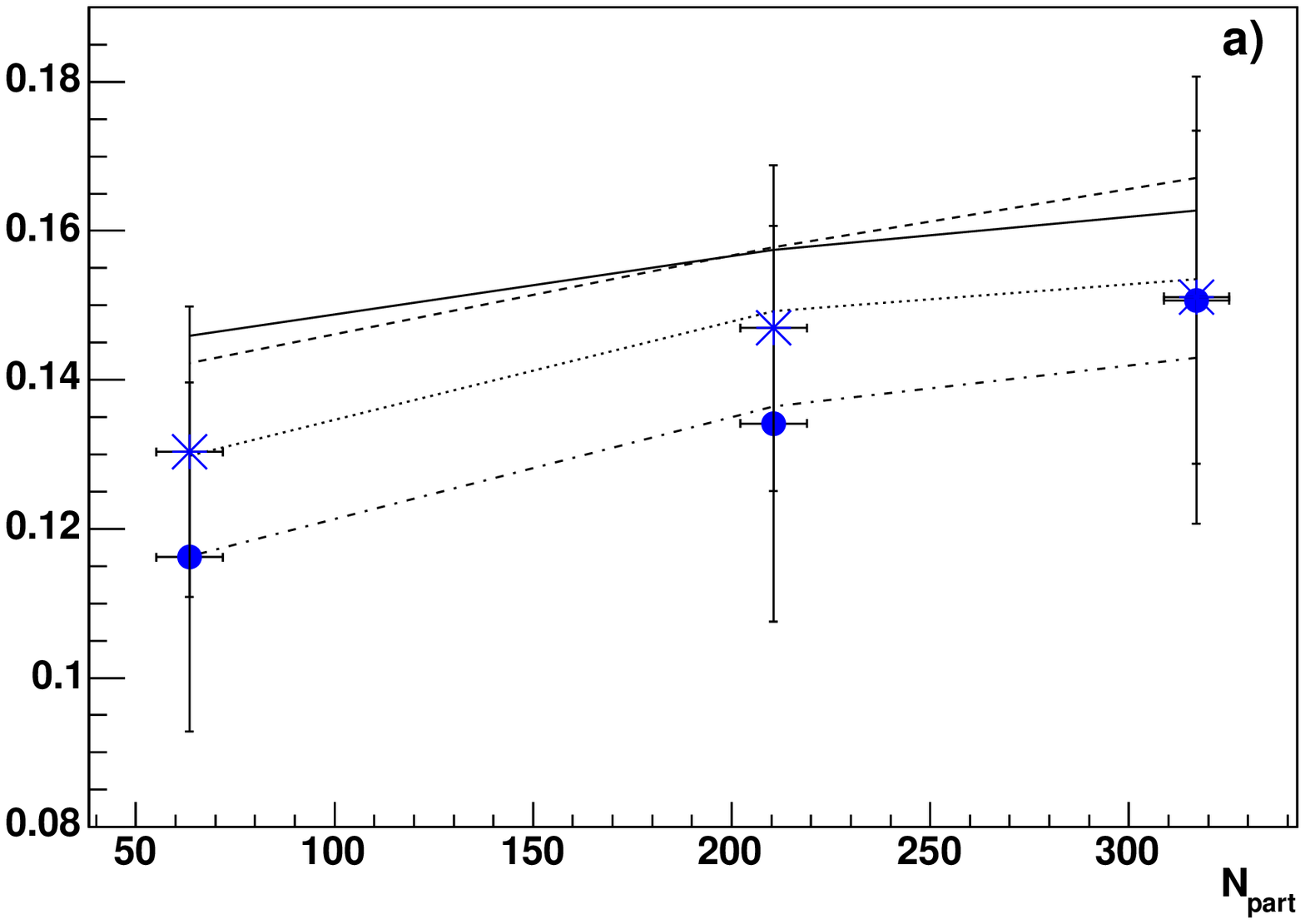}
\includegraphics[width=7cm]{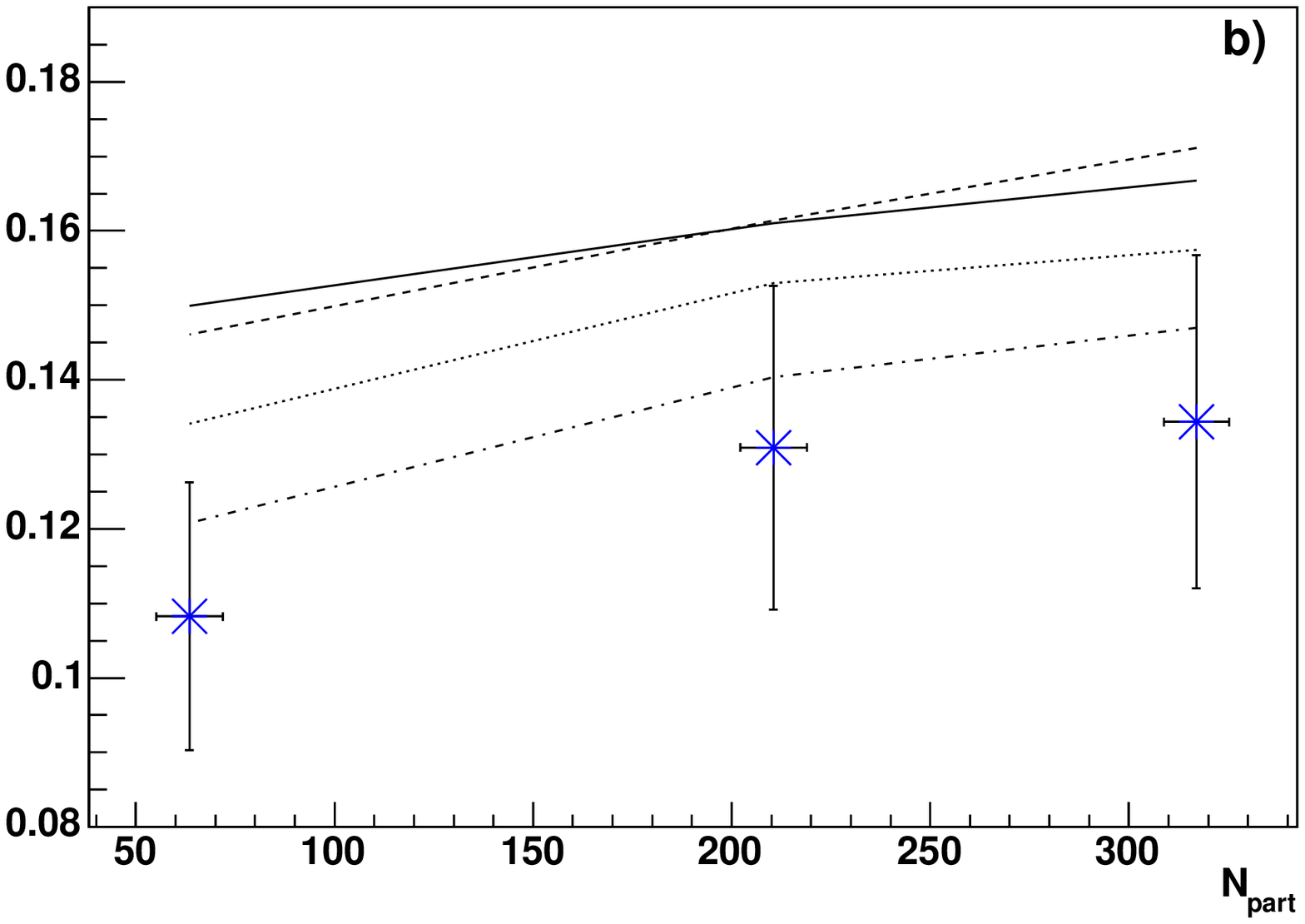}
\includegraphics[width=7cm]{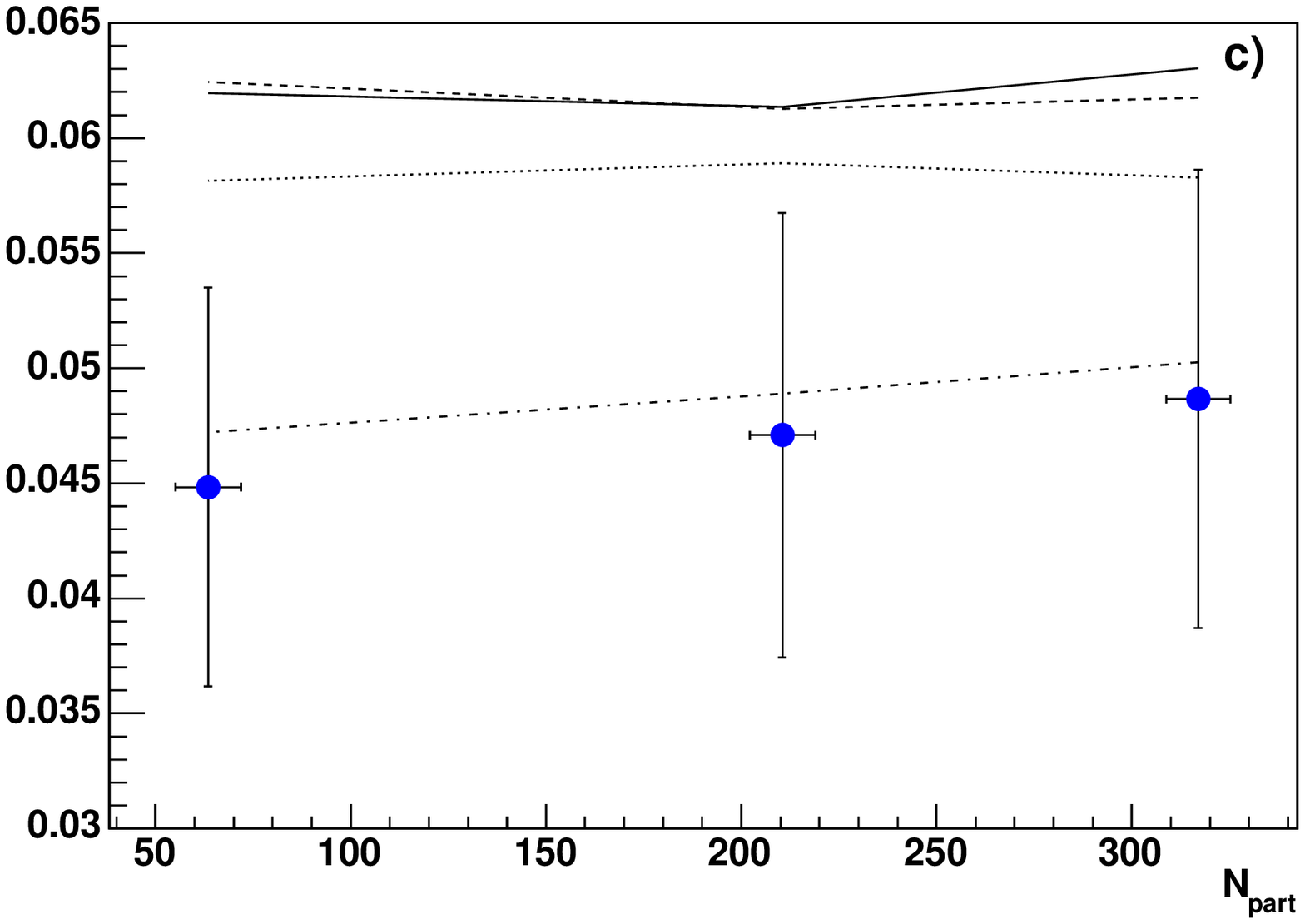}
\includegraphics[width=7cm]{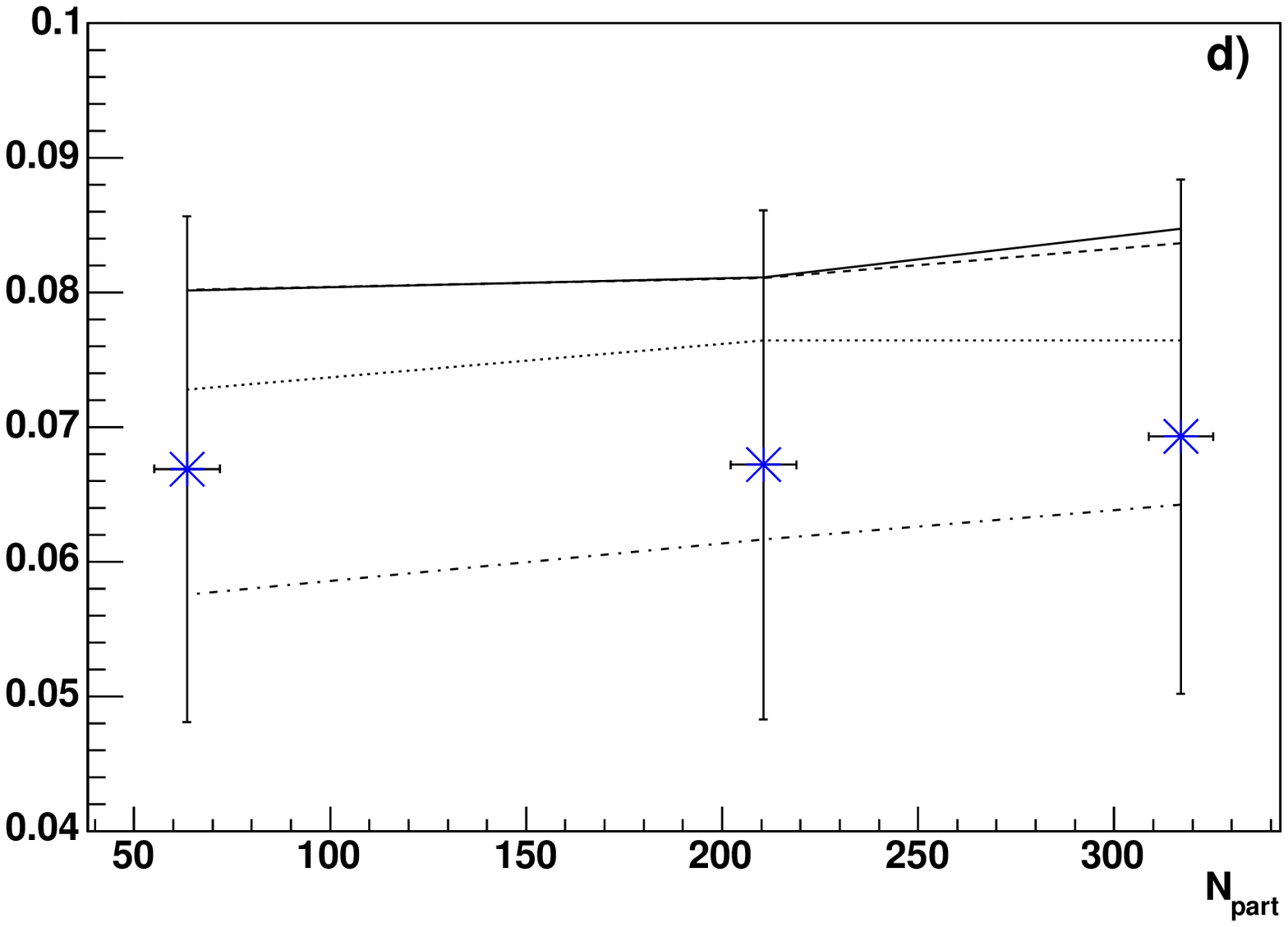}
\includegraphics[width=7cm]{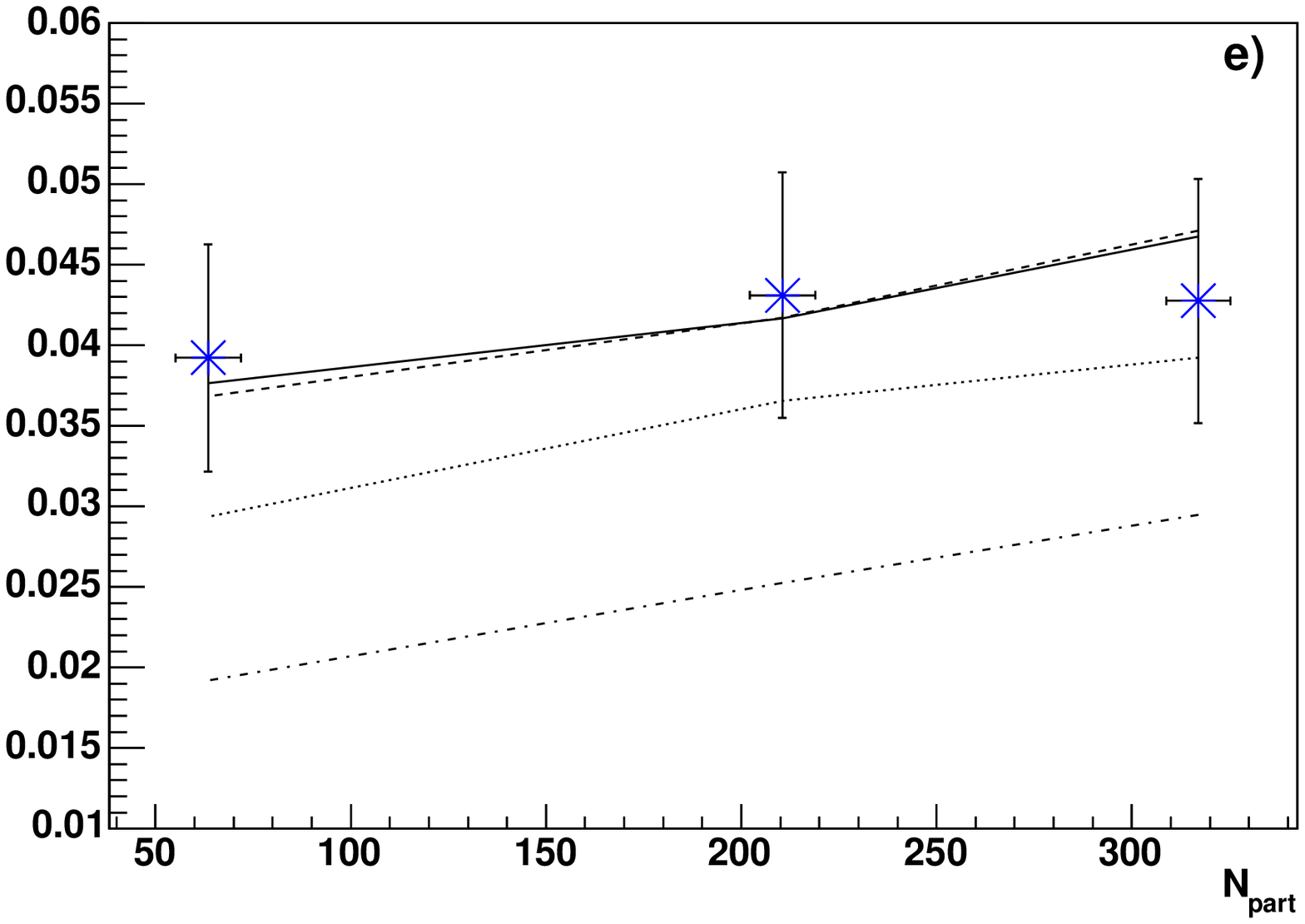}
\includegraphics[width=7cm]{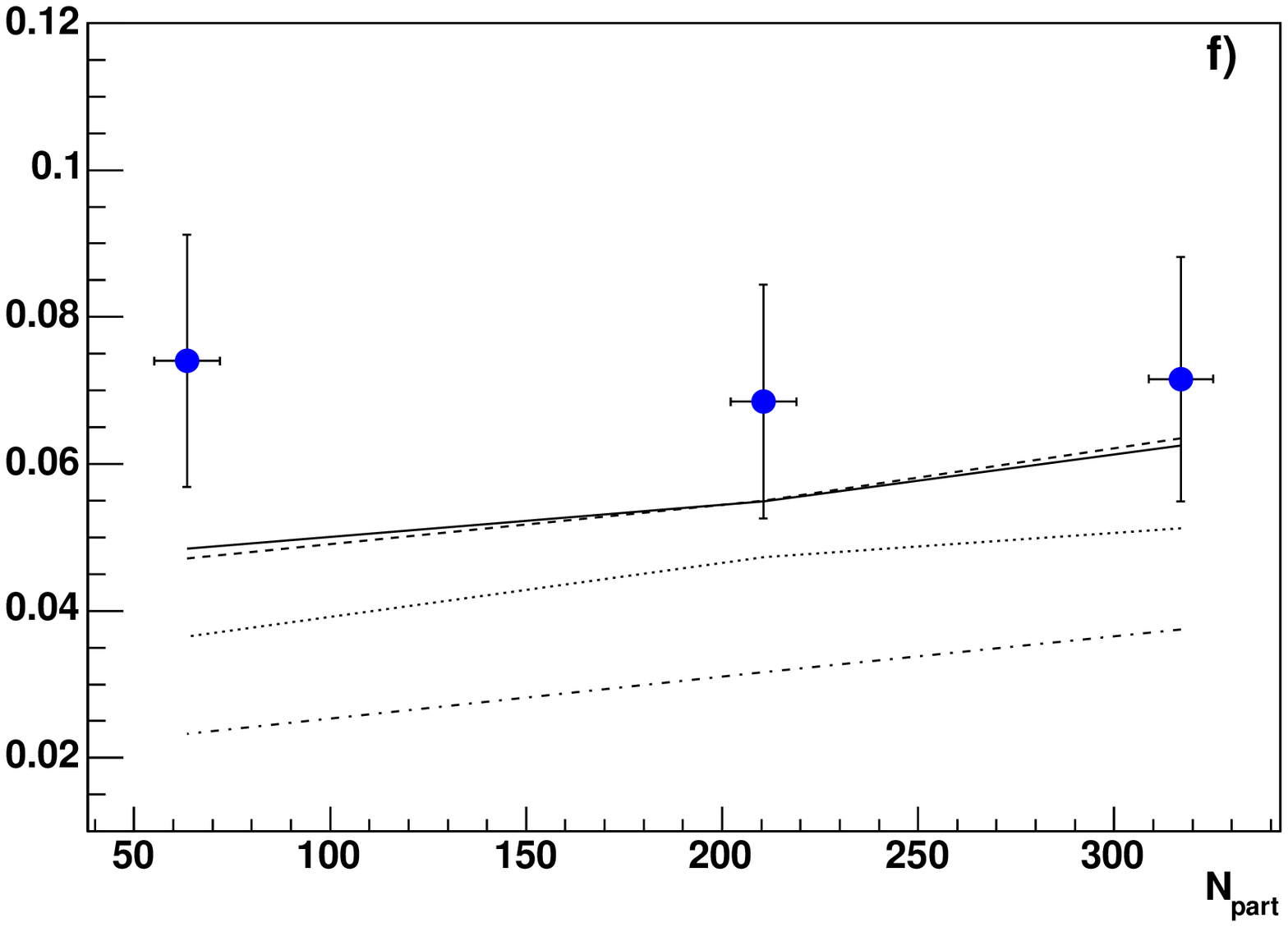}
\includegraphics[width=7cm]{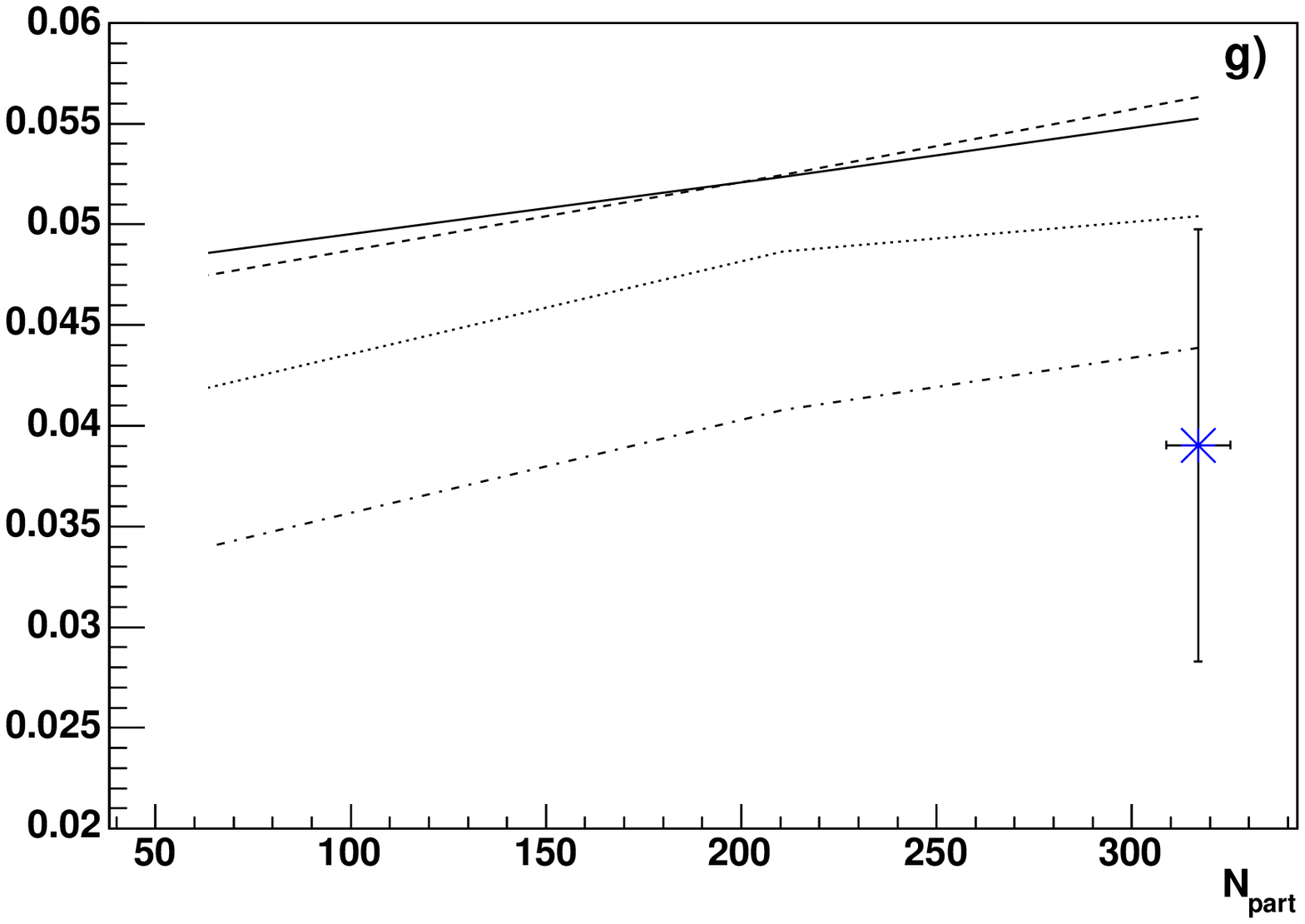}
\includegraphics[width=7cm]{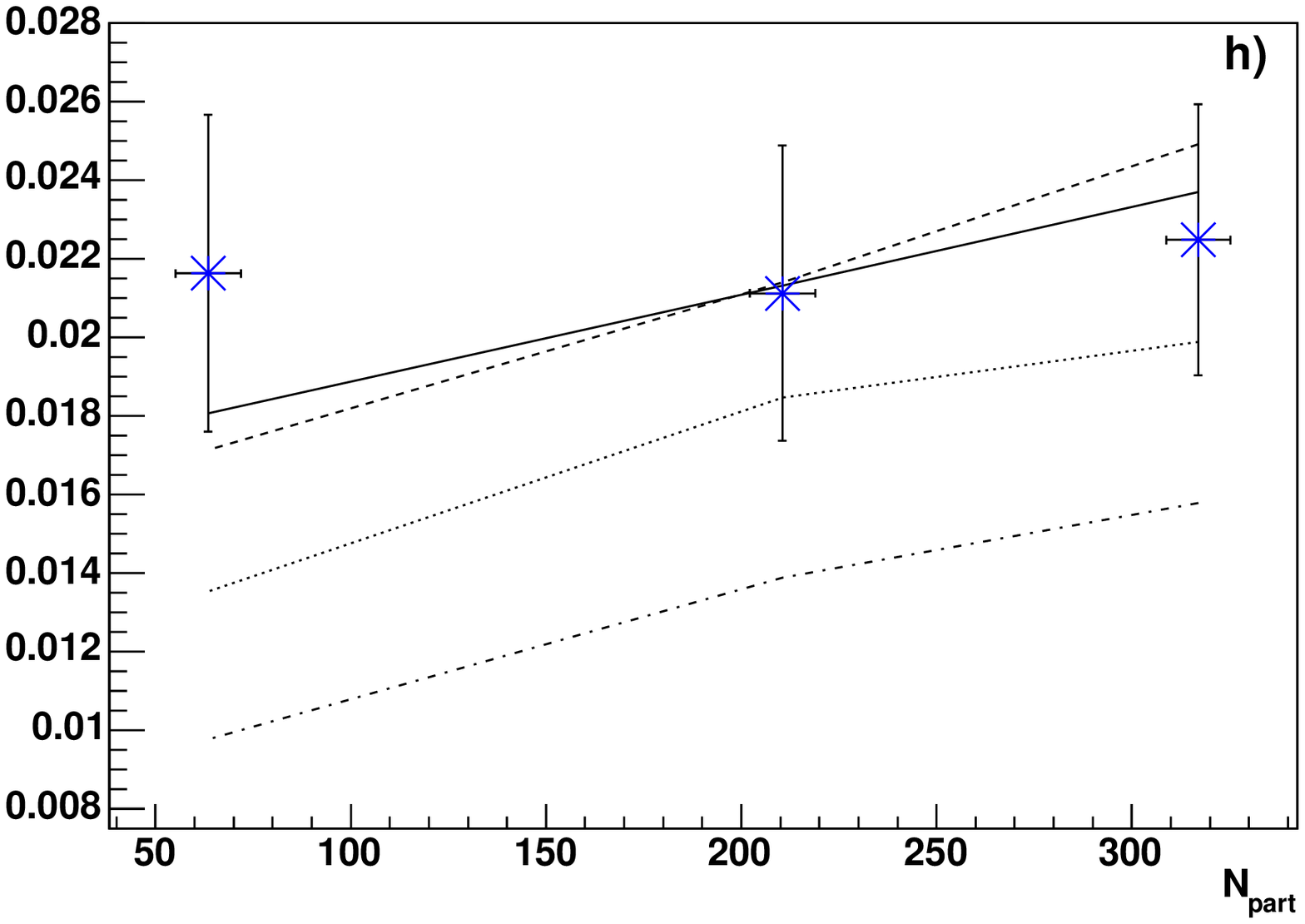}
\includegraphics[width=7cm]{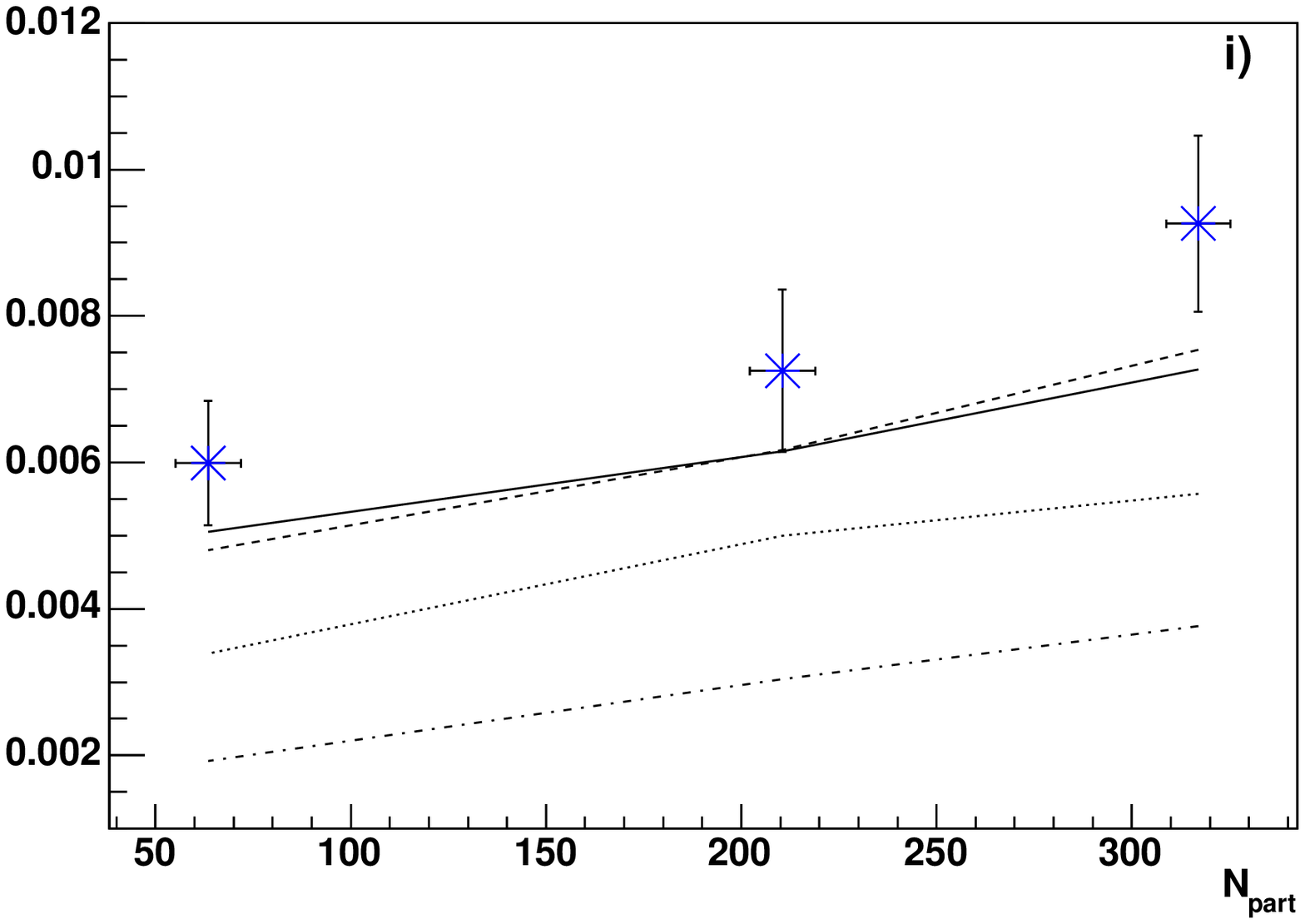}
\includegraphics[width=7cm]{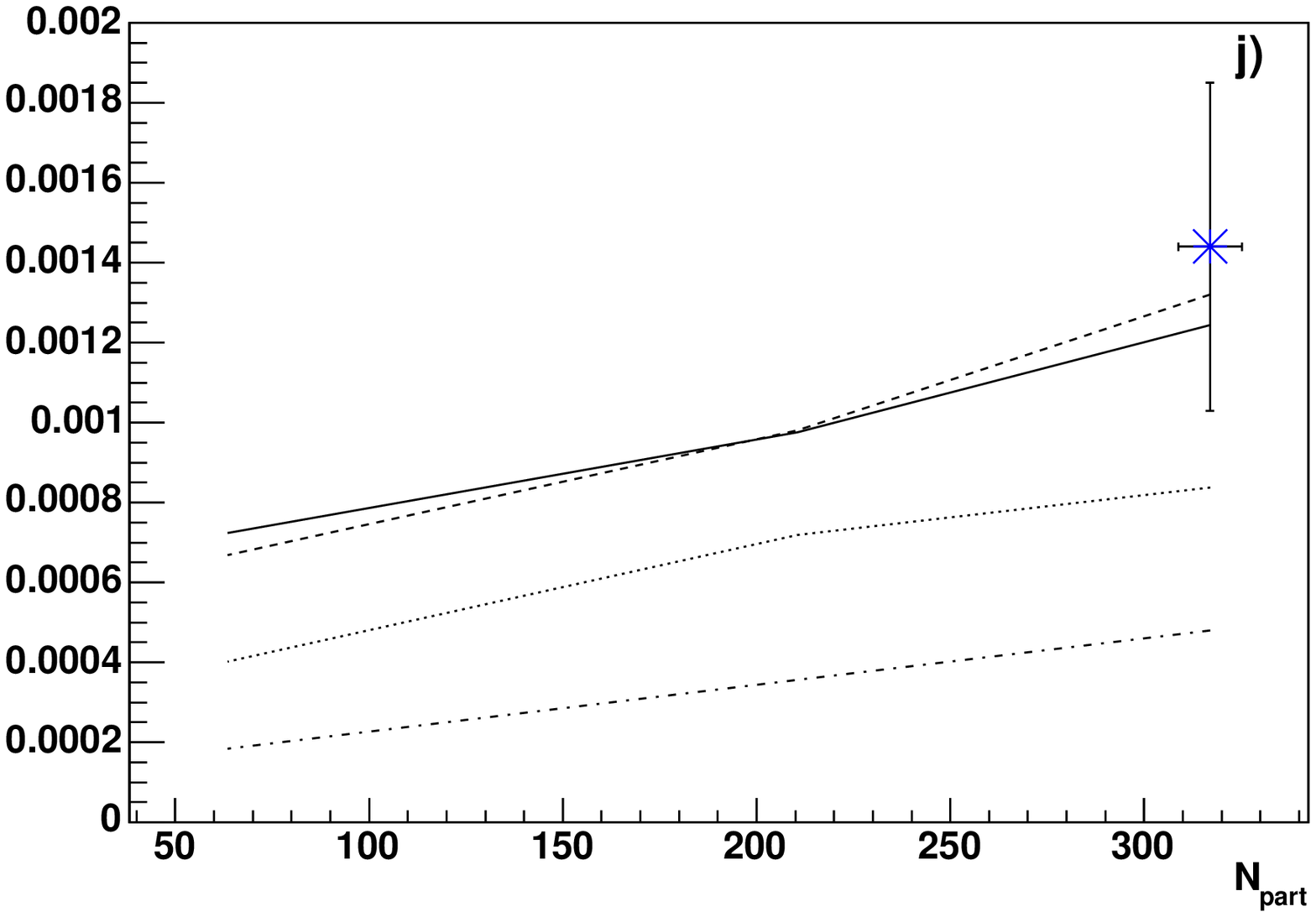}
\caption{As in Fig.~3 but for particle-to-pion ratios. 
(a: $K^-_{(2)}/\pi^-_{(2)}$, b: $K^0_{S}/\pi^-_{(2)}$, c: $\bar{p}_{(1)}/\pi^-_{(2)}$, 
d: $\bar{p}^-_{(2)}/\pi^-_{(2)}$, e: $\Lambda_{(1)}/\pi^-_{(2)}$, 
f: $\Lambda_{(2)}/\pi^-_{(2)}$, g: $<K^{*0}>/\pi^-_{(2)}$, 
h: $\phi/\pi^-_{(2)}$, i: $\Xi^-_{(2)}/\pi^-_{(2)}$, 
j: $\Omega^-/\pi^-_{(2)}$)}\label{Results:ModelPPi}
\end{figure}

For a qualitative understanding one can use the Boltzmann approximation 
ignoring widths. 
(At RHIC energies the use
of Boltzmann statistics introduces errors in primordial yields at the
level of 10\% for pions, 1.3-1.5\% for kaons and less than one percent
for all other hadrons. Thus, given the intrinsic systematic
uncertainties inherent in any thermal model analysis owing to the poorly
constrained particle properties of the heavy resonances, quantum
statistics is important only for pions.)
The primordial density of hadron species $i$ is then given by
\begin{equation}
n_i = {g_i\over(2\pi^2)}\gamma_S^{|S_i|} m_i^2 T K_2(m_i/T) \exp(\mu_i / T).
\end{equation}
Thus, the ratio of particle to anti-particle primordial yields is simply,
\begin{equation}
{n_i\over n_{\bar{i}}} = \exp(2\mu_i / T).
\end{equation}
Anti--particle-to-particle ratios therefore fix $\mu_{B,S}/T$.
We note that these ratios are only slightly affected by feed-down (using the 
central Fit I best-fit parameters the percentage
influence of feed-down is $\pi^-_{(2)}/\pi^+_{(2)}$: 1.0\%, $K^+_{(2)}/K^-_{(2)}$: 2.9\%,
$\bar{p}_{(1)}/p_{(1)}$: 0.5\%, $\bar{p}_{(2)}/p_{(2)}$: 3.7\%, 
$\bar{\Lambda}_{(1)}/\Lambda_{(1)}$: 0.7\%, 
$\bar{\Lambda}_{(2)}/\Lambda_{(2)}$: 4.2\%,
$\bar{\Xi}^+_{(2)}/\Xi^-_{(2)}$: 0.08\%, and $\bar{\Omega}/\Omega$: 0\%).
Thus, the trends in these ratios translate directly into trends in
$\mu_S/T$ and $\mu_B/T$, while they are roughly insensitive to
$\gamma_S$ and $T$.

Let us now consider the individual ratios and begin with $\pi^-_{(2)} / \pi^+_{(2)}$.
The experimental ratio is consistent with being flat and slightly below or equal to 
unity, while our model yields a $\pi^-/\pi^+$ ratio slightly above 1
(see Fig.~\ref{Results:ModelPAP}a).
Model and experiment, however, agree within errors. There is no significant
difference between the
results of the four fits. Setting $\mu_Q=0$ drives the model ratio to
very close to 1; deviations are due to feeding (since $\mu_B$ and
$\mu_S$ are both greater than zero). With only strong decays included, the
ratio is less than 1. Therefore it is the
influence of the weak decays that pushes the ratio above 1. This ratio
is insensitive to $\mu_S$ and
$\mu_B$: with a $5\ MeV$ change in $\mu_B$ the ratio varies by 0.2\%, while
a $5\ MeV$ change in $\mu_S$ leads to a 0.3\% change in the ratio. A 0.2
change in $\gamma_S$ furthermore affects the ratio by only 0.2\%,
while a $5\ MeV$ change in $T$ leads to only a 0.06\% change in the
ratio. Therefore, this ratio fixes nothing, and serves just as the
motivation for setting $\mu_Q=0$.

The ratio $K^+_{(2)} /K^-_{(2)}$ is above 1 (see Fig.~\ref{Results:ModelPAP}b) 
and drives $\mu_S$ to positive values. 
Due to their much smaller errors, the STAR ratios are heavily weighted in the fit. 
This contributes to $\mu_S$
having a slight kink as seen in Fig.~\ref{Results:Graph}c. 
The effect of the above-quoted feeding accounts for the slightly
different shape of the $K^+/K^-$ ratio compared with the $\mu_S/T$
trend which can be deduced from Fig.~\ref{Results:Graph}. 
With this ratio removed from the fit, $\mu_S/T$
would increase steeply with centrality, driven by the corrected 
$\bar{\Lambda}/\Lambda$ to $\bar{\Xi}^+/\Xi^-$ double ratio. This gives
$\bar{\Xi}^+/\Xi^-$ an increasing trend. In order to keep the
$\bar{\Lambda}/\Lambda$ ratio flat, $\mu_B/T$ then similarly increases
which causes $\bar{p}/p$ to drop. Therefore, the kaon ratio is what
keeps $\mu_S$ flattish and gives it the `kink'.

The $\bar{p}/p$ experimental ratios differ noticeably 
(see Fig.~\ref{Results:ModelPAP}c and Fig.~\ref{Results:ModelPAP}d), 
with STAR being the highest. These STAR data points are favored in fits since they have much
reduced errors (consistently a factor of 3 less than those of
PHENIX). These STAR ratios, however, are too high if the STAR $\bar{\Lambda}/\Lambda$ 
ratio is to be simultaneously reproduced by the
model (STAR $\bar{p}/p$ supports a lower $\mu_B$ as is seen
when the $\Lambda$'s are removed from the fit, as done in Fit IV).
With $\Lambda$'s excluded, the much smaller STAR errors lead to these
$\bar{p}/p$ data points being strongly favored in the mid-central and
peripheral fits. This leads to a strong increase of $\mu_B$ with centrality. Since
the experimental
$\bar{\Lambda}/\Lambda$ ratio (Fig.~\ref{Results:ModelPAP}e and f) 
is flattish, they
support a similar behavior for $\mu_B/T$ and $\mu_S/T$.

Excluding all but the STAR $\bar{p}/p$ points in the most central bin
removes the strong increase in $\mu_B$ from the mid-central to central bins
seen in Fit IV. Our suspicion that it is $\bar{p}/p$ which
causes $\mu_B$ to increase is confirmed by repeating the
fits with only anti--particle-to-particle ratios included
(i.e., all particle-to-pion ratios excluded). Since these ratios are insensitive
to $\gamma_S$, and $T$ enters only together with the chemical
potentials, in these reduced fits we fixed $\gamma_S=1$ and $T=165\ MeV$
(both reasonable in light of the full results). The approximate trends in
$\mu_S/T$ and $\mu_B/T$ observed in the full fits were reproduced
using just these ratios. With the
$\bar{p}/p$ ratio excluded, the steady increase in $\mu_B$ disappears;
instead $\mu_B/T$ decreases similarly to $\mu_S/T$ as driven by the
flatness in the $\bar{\Lambda}/\Lambda$ ratio.

The ratio $\bar\Xi^+ / \Xi^-$ (see Fig.~\ref{Results:ModelPAP}g) 
depends strongly on $\mu_S/T$ and $\mu_B/T$ and is little
affected by feed-down. Similar fractional errors to the uncorrected 
$\bar{\Lambda}/\Lambda$ ratio means both are treated equally in fits.

Although consistent within errors, the model trend decreases slightly
with centrality, while the data are consistent with a weak increase. Comparing
this experimental ratio with $\bar{\Lambda}/\Lambda$, the 
$\bar{\Xi}^+/\Xi^-$ to $\bar{\Lambda}/\Lambda$ double ratio increases. Thus, 
$\mu_S/T$ is driven up particularly from the mid-central
to central bin. As mentioned earlier, the $K^+/K^-$ ratio leads to a
kink in the trend for $\mu_S$. This effect wins over the influence of the $\Xi$ to
$\Lambda$ double ratio which suggests an increase in $\mu_S/T$.

Unfortunately there is no centrality information on the ratio
$\bar{\Omega}^+ / \Omega^-$. 
Such data would allow even better determination of $\mu_S$. 
As seen in Fig.~\ref{Results:ModelPAP}h, 
our model results
agree fairly well with the data for the central bin.

\subsubsection{Particle-to-pion ratios}

The particle-to-pion ratios read in the Boltzmann approximation neglecting width
\begin{equation}
{n_i \over n_\pi} = \gamma_S^{|S_i|} {g_i \over g_{\pi}}
{m_i^2 \over m_\pi^2}
{K_2(m_i/T) \over K_2(m_\pi/T)}
{\exp(\mu_i / T) \over \exp(\mu_\pi / T)},
\end{equation}
i.e., qualitatively, mixed particle ratios allow the determination of
$\gamma_S$ and $T$, and hence the $\mu$'s independently.
Whereas the particle-to-anti--particle ratios feature
just the combination
$\mu/T$, the primordial particle-to-pion ratios feature also $T$ in the
combination $K_2(m_i/T) / K_2(m_{\pi}/T)$. This allows the temperature to
be fixed and then, given the values of $\mu/T$ fixed by the
particle-to-anti-particle ratios, the chemical potentials are
determined. However, feed-down plays an important role in these
ratios, so that the final observed ratios differ from the
ratios of primordial densities quite substantially. This makes a
detailed analysis of the influence
of these ratios on the parameters much more difficult than in the case of the
particle-to-anti--particle ratios. In fact, for the best-fit parameters 
corresponding to Fit I of the central bin, more than 70\%
of $\pi^+$'s come from decays.

A simple analysis shows how the ratio $K_2(m_i/T) / K_2(m_{\pi}/T)$ 
varies with temperature for a number of hadron species $i$:
Lowering the temperature has the
effect of bringing the particle-to-pion ratios down. Furthermore, the
fractional effect of changing the temperature is greater for the
heavier particles.

After these general remarks let us consider in some detail the 
comparison of our model with data.
The ratio ${K^-_{(2)} / \pi^-_{(2)}}$ (Fig.~\ref{Results:ModelPPi}a)
is reasonably well reproduced in the fits.
Since the errors on the PHENIX and STAR experimental ratios are similar,
neither is heavily biased in the fit. The PHENIX data rise more rapidly than
the STAR data (especially from the mid-central to central bin). Since
the particle-to-anti-particle ratios support a centrality-independent
$\mu_S/T$, the experimental $K^-/\pi^-$ ratios drive an increase in strangeness
saturation, with the PHENIX data supporting a steeper increase in $\gamma_S$ 
than the STAR data (see Fig.~\ref{Results:Graph}d).
In Fit I and II $K^-/\pi^-$ is 
over-predicted. In contrast, the uncorrected $\Lambda/\pi^-$ and $\Xi^-/\pi^-$ 
are under-estimated (small fractional errors on $\Xi^-/\pi^-$ mean that this 
ratio strongly influences the fit). With the $\Xi$'s removed in Fit III the STAR $K^-/\pi^-$ 
data are well reproduced owing to the drop in $T$ and $\gamma_S$.
The big drop in this ratio when $\Lambda$'s are excluded from the fit
is due to the further drop in $\gamma_S$ and $T$ in Fit IV seen in Fig.~\ref{Results:Graph}.
The effect of this drop is somewhat lessened by
the accompanying drop in $\mu_S/T$ in the central and mid-central bins. Furthermore, 
the kaon to pion ratios are
least affected by the temperature change in Fit IV.

The increase in the ratio ${K^0_S / \pi^-}$ with centrality
(Fig.~\ref{Results:ModelPPi}b) also drives $\gamma_S$ to increase. In the full
fits (Fits I - III) it is somewhat over-estimated (a similar result is found in
\cite{NuXuKaneta}), while the exclusion of the $\Lambda$'s (Fit IV)
results in much better agreement. This is owing to the reduced
$\gamma_S$ in Fit IV. Comparing the results for the ratios $K^- / \pi^-$ and 
$K^0_S / \pi^-$ in Figs.~\ref{Results:ModelPPi}a and b one gets the impression
that the PHENIX $K^- / \pi^-$ data is compatible with STAR's $K^0_S / \pi^-$
data, while the good fit in Fit III of STAR's $K^- / \pi^-$ data result
in a ratio $K^0_S / \pi^-$ being at the upper limits of the error bars from STAR. 
The central $<K^{*0}>/\pi^-$ ratio reproduction by the model (Fig.~\ref{Results:ModelPPi}g) 
follows the same trend as the $K^0_S$.

Let us now focus on the $\bar p / \pi^-$ and $\Lambda / \pi^-$ 
ratios (Figs.~\ref{Results:ModelPPi}c, d, e and f). 
While the $\bar{p}/\pi^-$ ratio is over-estimated by the model in Fits I - III, 
$\Lambda/\pi^-$ is under-estimated except for the corrected $\Lambda/\pi^-$ in 
the central bin in Fits I and II. With $\Lambda$ excluded, 
$\bar{p}/\pi^-$ is well reproduced, 
while $\Lambda/\pi^-$ is greatly underpredicted.
Therefore, $\bar{p}/\pi^-$ drives $T$ down, while $\Lambda/\pi^-$ tends
to raise $\gamma_S$ and $T$.

The flatness of the ratio $\phi / \pi^-$ (Fig.~\ref{Results:ModelPPi}h)
supports a centrality-independent $\gamma_S$. In Fits I and II there is fair agreement with
the model, while this ratio is greatly under-estimated in Fit IV owing 
to the greatly reduced $\gamma_S$ and $T$. The increasing trend of the
model results is due to the increase in $\gamma_S$ with
centrality and the fact that $n_{\phi}\sim\gamma_S^2$.

The increase of the ratio $\Xi^- / \pi^-$ (Fig.~\ref{Results:ModelPPi}i)
drives $\mu_B/T$ to increase as well as
$\gamma_S$. On the other hand, it causes $\mu_S/T$ to drop. Again there is fair agreement
with the model in Fits I and II, while this ratio is greatly under-estimated in Fit IV.

There is fair agreement of our model results with the $\Omega^-/\pi^-$ ratio 
(Fig.~\ref{Results:ModelPPi}j) in 
the central bin in Fits I and II. However, with $\Xi$ and $\Lambda$ removed 
agreement worsens.

Our final conclusion is that Fits I - II give a quite satisfactory description
of most ratios, with the exception of $K^0_S / \pi^-$, $<K^{*0}>/\pi^-$ and corrected 
$\bar p / \pi^-$. For the $<K^{*0}>/\pi^-$ and $\Omega^- / \pi^-$ there are no data for 
mid-central and peripheral collisions. Thus our results may serve as predictions.

\subsection{Summary}

In summary, we present an analysis of the centrality dependence
of thermal parameters deduced from hadron multiplicities within a 
statistical-thermal model. 
The grand-canonical formalism is employed with quantum statistics and
resonance widths included for all of the fits. 
We rely entirely on published data for the reaction
Au + Au at $\sqrt{s_{NN}} = 130\ GeV$. Our aim is to employ the strangeness saturation factor
$\gamma_s$ as one possible indicator for deviations from equilibrium conditions
in describing the chemical freeze-out. Indeed, for non-central collisions
the systematic deviation of $\gamma_s$ from unity can be considered as a hint to
off-equilibrium effects, while for central collisions and using a sufficiently
complete set of available multiplicity ratios at mid-rapidity
there is no indication for deviations from equilibrium conditions.

We present a detailed comparison of the model results with data.
The trends in $\mu_{B,S} / T$ are fixed by the
anti--particle-to-particle ratios. The $K^+/K^-$ ratio is responsible
for keeping $\mu_S / T$ essentially flat. This wins over the
influence of the $\Xi$ to
$\Lambda$ double ratio which tends to increase $\mu_S/T$. The
$\bar{p}/p$ ratio drives $\mu_B/T$ up, which
wins over $\bar{\Lambda}/\Lambda$'s attempts to make $\mu_S/T$ and
$\mu_B/T$ behave similarly.
The temperature and strangeness saturation are fixed by the
particle-to-pion ratios. The observed increase in $\gamma_S$ is driven predominantly
by the increase in the $K^-/\pi^-$ ratio and the flatness of $\mu_S/T$ as fixed by
the $K^+/K^-$ ratio. 

A possible extension of the present work is the analysis of saturation in
the non-strange sector, parameterized by $\gamma_q$ being a factor similar
to $\gamma_s$ \cite{Rafelski99}. Since our fits to the data deliver reasonably
small values of $\chi^2$ we believe, however, that there is not too much room
left for sizeable deviations of $\gamma_q$ from unity. This issue deserves
a separate investigation. More tempting is the analysis of data for
$\sqrt{s_{NN}} = 200\ GeV$. The explorative study in \cite{NuXuKaneta}
points to a similar behavior of $\gamma_s$ as we find for
$\sqrt{s_{NN}} = 130\ GeV$.
Even more interesting is to test heavy flavor production; due to their
larger intrinsic masses, heavy flavors should have a higher sensitivity to
the medium. In particular, hadrons with open and hidden charm may require
a corresponding factor $\gamma_c$.

\subsection*{Acknowledgments}

This work has been supported in part by the National Research Foundation of
South Africa, the U.S. Department of Energy under
Contract No.\ DE-AC03-76SF00098 and the German BMBF grant 06DR121.

\end{document}